\documentclass[twoside,leqno,twocolumn]{article}
\pdfoutput=1 
\usepackage[letterpaper]{geometry}
\usepackage{ltexpprt}

\usepackage{color}
\usepackage[utf8]{inputenc}
\usepackage{longtable}

\usepackage{xspace}
\usepackage{appendix}
\usepackage{booktabs}
\usepackage[utf8]{inputenc}
\usepackage{subfig}
\usepackage[font={small,bf},labelfont=bf,format=plain]{caption}
\usepackage{textcomp}
\usepackage{amsmath, amssymb}
\usepackage{enumitem}
\usepackage{algorithm}
\usepackage{algpseudocode}
\algtext*{EndProcedure}
\usepackage{xcolor,colortbl}
\usepackage{footnote}
\usepackage{url}
\usepackage{tabularx}
\usepackage{multicol}
\usepackage{multirow}
\usepackage{pifont}
\usepackage{bm}
\usepackage{listings}
\usepackage{wrapfig}
\usepackage{relsize}
\usepackage{mathtools}
\usepackage{bbm}
\usepackage{blindtext}
\usepackage{bm}
\usepackage{balance}
\usepackage{mdwlist}
\setlist[enumerate]{leftmargin=*}
\setlist[itemize]{leftmargin=*}

\usepackage{wasysym}
\usepackage{amssymb}
\usepackage{pifont}

\newcommand{\adj}{\mathbf{A}}
\newcommand{\graph}{G}
\newcommand{\vertexSet}{\mathcal{V}}
\newcommand{\numberOfAllNodes}{n}
\newcommand{\numberOfNodes}[1]{n_{#1}}
\newcommand{\degavg}{\bar{d}}
\newcommand{\degmax}{d_{\text{max}}}
\newcommand{\edgeSet}{\mathcal{E}}
\newcommand{\numTop}{k}

\newcommand{\matchMat}{\mathbf{M}}
\newcommand{\matA}{\mathbf{A}}
\newcommand{\updateMatch}{\mathbf{U}}

\newtheorem{problem}{Problem}
\newtheorem{observation}{Observation}
\newtheorem{definition}{Definition}

\newcommand{\method}{\text{RefiNA}\xspace}

\newcommand{\lr}{\eta}
\newcommand{\reg}{\lambda}

\newcommand{\insightone}{\texttt{I1}\xspace}
\newcommand{\insighttwo}{\texttt{I2}\xspace}
\newcommand{\insightthree}{\texttt{I3}\xspace}

\begin{document}

\title{Refining Network Alignment \\ to Improve Matched Neighborhood Consistency}

\author{Mark Heimann\thanks{Computer Science \& Engineering, University of Michigan. Email: \{mheimann, shinech, vfatemeh, dkoutra\}@umich.edu} \footnote{Now at Lawrence Livermore National Laboratory.}\\
\and
Xiyuan Chen\footnotemark[1]~\footnote{Now at Stanford University.}\\
\and
Fatemeh Vahedian\footnotemark[1]\\
\and
Danai Koutra\footnotemark[1] 
}
\date{}

\maketitle

\fancyfoot[R]{\scriptsize{Copyright \textcopyright\ 2021 by SIAM\\
Unauthorized reproduction of this article is prohibited}}

\begin{abstract}
Network alignment, or the task of finding meaningful node correspondences between nodes in different graphs, is an important graph mining task with many scientific and industrial applications.  
An important principle for network alignment is \emph{matched neighborhood consistency} (MNC): nodes that are close in one graph should be matched to nodes that are close in the other graph.  We theoretically demonstrate a close relationship between MNC and alignment accuracy.  
As many existing network alignment methods struggle to preserve topological consistency in difficult scenarios, we show how to refine their solutions by improving their MNC. 
Our refinement method, \method, is straightforward to implement, admits scalable sparse approximation, and can be paired \emph{post hoc} with \emph{any} network alignment method.  Extensive experiments show that \method increases the accuracy of diverse unsupervised network alignment methods by up to 90\%, making them robust enough to align graphs that are $5\times$ more topologically different than were considered in prior work.

\end{abstract}

\section{Introduction}
\label{sec:intro}
Network alignment, or the task of finding correspondences between the nodes of multiple networks, is a fundamental graph mining task with applications to user identity linkage~\cite{liu2016aligning}, discovery of novel biological functions~\cite{graal}, computer vision~\cite{gold1996graduated}, schema matching in databases~\cite{melnik2002similarity}, and other problems of 
academic and industrial interest.  
Methods to solve this problem vary widely in techniques, ranging from nonlinear relaxations of a computationally hard optimization problem~\cite{aflalo2015convex}, 
to belief propagation~\cite{netalign}, 
genetic algorithms~\cite{magna}, 
spectral methods~\cite{isorank}, 
node embedding~\cite{regal}, and more.  

Matching the topological structure of graphs is difficult, closely related to the canonical graph isomorphism problem~\cite{aflalo2015convex}.  
As a result, many network alignment approaches 
rely heavily on node or edge side information~\cite{final,meng2016igloo} (in some cases ignoring the graph structure altogether),
or on known ground-truth alignments---used as anchor links to connect the two graphs~\cite{liu2016aligning, zhang2019origin}
or to supervise the training of deep neural networks~\cite{dgmc}. 
However, 
in many settings~\cite{regal,dana,du2019joint}, side information or anchor links may not be available. 

\begin{figure}
    \centering
    \includegraphics[width=0.48\textwidth]{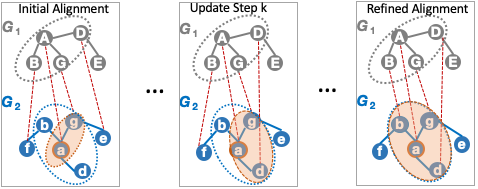}
    \caption{\method refines an initial network alignment solution, which maps node A and its neighbors in $\graph_1$ far apart in $\graph_2$. The refined network alignment solution has higher \emph{matched neighborhood consistency}: neighbors of A are aligned to neighbors of a, to which A itself is aligned.}
    \label{fig:method}
    \vspace{-0.3cm}
\end{figure}

In this paper, we focus on the challenging problem of unsupervised topological network alignment.  With neither anchor links to seed the alignment process nor side information to guide it, the main objective for this task is to preserve some kind of topological consistency in the alignment solution.  We theoretically analyze the principle of \emph{matched neighborhood consistency} (MNC), or how well a node's neighborhood maps onto the neighborhood of its counterpart in the other graph (illustrated in Fig.~\ref{fig:method}), and show its connection to alignment accuracy.
On the other hand, we find that when network alignment methods are inaccurate, the MNC of their solutions breaks down (e.g., Fig.~\ref{fig:method} left). 

To address this, we introduce \method, 
a method for refining network alignment solutions \emph{post hoc} by iteratively updating nodes' correspondences to improve their MNC.  By strategically limiting the possible correspondences per node to update in each iteration, we can sparsify the computations to make \method scalable to large graphs. 
Experimentally, 
we show that \method significantly improves a variety of network alignment methods on many highly challenging datasets, even when starting a network alignment solution with limited accuracy.  
Also, it can be succinctly expressed as matrix operations in a few lines of code, making it easy for practitioners to adopt.  In this compact formulation, we incorporate several useful insights for network alignment, and our techniques have interesting connections to other successful graph-based methods.

Our contributions are as follows:
\begin{itemize}
    \item \textbf{New Algorithm}: We propose \method, a post-processing step that can be applied to the output of any network alignment method.  Its compact design incorporates several important insights for network alignment, and it permits a sparse approximation that is scalable to large graphs.
    \item \textbf{Theoretical Connections}: We show a rigorous connection between \emph{matched neighborhood consistency}, the property that \method improves, and alignment accuracy.  We also provide network alignment insights justifying each of \method's design choices and technical connections to other graph-based methods. 
    \item \textbf{Experiments}: We conduct thorough experiments on real and simulated network alignment tasks and show that \method improves the accuracy of many methodologically diverse network alignment methods by up to 90\%, making them robust enough to recover matchings in $5\times$ noisier datasets than those considered in prior work.  We extensively drill down \method to justify the insights that inspire its design.
\end{itemize}
We provide our code and additional supplementary material at \url{https://github.com/GemsLab/RefiNA}.

\section{Related Work}
\label{sec:related}

Network alignment 
has numerous applications from matching schemas in databases~\cite{melnik2002similarity} and objects in images~\cite{gold1996graduated}, to revealing biological functions shared by different organisms~\cite{graal}, to integration of multiple data sources to create a holistic {worldview network}~\cite{douglas2018metrics}.  as been widely studied and several methods have been proposed to solve this problem. 
For example, the intuition of NetAlign~\cite{netalign} as a message-passing algorithm is to ``complete squares'' by aligning two nodes that share an edge in one graph to two nodes that share an edge in another graph.  Similarly, FINAL~\cite{final}
has an objective of preserving topological consistency between the graphs that may be augmented with node and edge attribute information, if available.  MAGNA~\cite{magna} is a genetic algorithm that can evolve network populations to maximize topological consistency. Recent works leverage kernel methods~\cite{zhang2019kergm} or optimal transport~\cite{xu2019gromov} but suffer from high (cubic) computational complexity. 

Networks can also be aligned by comparing nodes directly.  Early works hand-engineered node features: GRAAL \cite{graal} computes a graphlet degree signature, while GHOST~\cite{patro2012global} defines a multiscale spectral signature. 
UniAlign~\cite{bigalign} and HashAlign~\cite{hashalign} extract features for each node from graph statistics such as degree and various node centralities. 
Recent works instead leverage node embeddings that are very expressive, but must be comparable across networks~\cite{regal}.  REGAL~\cite{regal} computes node embeddings that capture structural roles that are not specific to a particular network.  Other methods align the embedding spaces of different networks using techniques from unsupervised machine translation: DANA~\cite{dana} uses adversarial training~\cite{lample2018word}, and CONE-Align~\cite{conealign} uses non-convex alternating optimization methods.  All these approaches find match nodes with embedding-based nearest-neighbor search, which does not enforce alignment consistency and can benefit from our refinement.  Our contributions are orthogonal to a recent effort to accelerate embedding-based node matching via graph compression~\cite{qin2020g}.

Some graph neural matching network models, often in specific applications like social networks~\cite{li2019adversarial}, computer vision~\cite{dgmc}, or knowledge graphs~\cite{wu2020neighborhood}, use objectives that enforce matching consistency between neighboring nodes.  This is similar to our approach \method; indeed, in the supplementary \S\ref{app:connections} we show interesting technical connections between \method and graph neural networks.  However, \method, like Simple Graph Convolution~\cite{sgc} is much faster and simpler to implement and apply than these graph neural networks, and it does not need known node matchings to supervise training.

\section{Theoretical Analysis}

We first introduce key alignment concepts and notation. Then, we theoretically justify the topological consistency principle that is the basis of our refinement approach, \method (\S\ref{sec:method}).  

\subsection{Preliminaries}
\label{sec:preliminaries}
\begin{table}[t!]
     \caption{Major symbols and definitions.}
     \label{tab:dfn}
     \vspace{-0.2cm}
{\footnotesize
\begin{tabular}{ p{1.8cm} p{6cm}}
     \toprule
     \textbf{Symbols} &  \textbf{Definitions}\\
     \midrule
     $\graph_\ell = (\vertexSet_\ell, \edgeSet_\ell)$ & $\ell^{th}$ graph  with nodeset $\vertexSet_\ell$, edgeset $\edgeSet_\ell$ \\
     $\adj_\ell$ & Adjacency matrix of $\graph_\ell$ \\
     $\numberOfNodes{\ell}, \degavg^\ell$ & Number of nodes and average degree in $\graph_\ell$ \\
     $\pi(\cdot)$ & An alignment between graphs $\graph_1$ and $\graph_2$; a function mapping a node in $\vertexSet_1$ to a node in $\vertexSet_2$ \\
     $\matchMat$ & $\numberOfNodes{1} \times \numberOfNodes{2}$ matrix specifying correspondences of nodes in $\vertexSet_1$ to those in $\vertexSet_2$ \\
     $\mathcal{N}_{\graph_\ell}(u)$ & Neighbors of node $i$ in graph $\graph_\ell$ \\
     $\tilde{\mathcal{N}}_{\graph_2}^\pi(u)$ & ``Mapped neighborhood'' of node $i$; counterparts in $\graph_2$ (mapped by $\pi$) of nodes in $\mathcal{N}_{\graph_1}(i)$ \\ 
     
     \bottomrule
    \end{tabular}
    }
    \end{table}
    \setlength{\textfloatsep}{1\baselineskip plus 0.2\baselineskip minus 0.5\baselineskip}

\subsubsection{Graphs.} Following the network alignment literature~\cite{regal,magna}, we consider two unweighted and undirected {graphs} $\graph_1 = (\vertexSet_1, \edgeSet_1)$ and $\graph_2 = (\vertexSet_2, \edgeSet_2)$  with their corresponding nodesets $\vertexSet_1,\vertexSet_2$ and edgesets $\edgeSet_1, \edgeSet_2$.  We denote their adjacency matrices as $\matA_1$ and $\matA_2$. Since they are symmetric, $\matA_1^\top=\matA_1$ and $\matA_2^\top=\matA_2$, and we simplify our notation below.

\subsubsection{Alignment.} 
A node alignment is a function $\pi: \vertexSet_1 \rightarrow \vertexSet_2$ that maps the nodes of $\graph_1$ to those of $\graph_2$. 
It is also commonly represented as  
a $|\vertexSet_1| \times |\vertexSet_2|$ alignment matrix $\matchMat$, where $\matchMat_{ij}$ is the (real-valued or binary) similarity between node $i$ in $G_1$ and node $j$ in $G_2$. 
$\matchMat$ can be used to encode a mapping $\pi$, e.g., greedy alignment $\pi(i) = \arg\max_j \matchMat_{ij}$.  
We note that alignment between two graphs should be sought \textit{if} the nodes of the two graphs meaningfully correspond.

\subsubsection{Neighborhood and Consistency.} Let $\mathcal{N}_{\graph_1}(i)=\{j\in\vertexSet_1: (i,j)\in\edgeSet_1\}$ be the neighbors of node $i$ in $\graph_1$, i.e., the set of all nodes with which $i$ shares an edge.  
We define node $i$'s ``\textbf{mapped neighborhood}'' in $\graph_2$ as the 
set of nodes onto which $\pi$ maps $i$'s neighbors: $\tilde{\mathcal{N}}_{\graph_2}^\pi(i) = \{j \in \vertexSet_2 : \exists k \in \mathcal{N}_{\graph_1}(i) \text{ s.t. } \pi(k) = j\}$.  For example, in Fig.~\ref{fig:method} (first panel), node $A$'s neighbors in $\graph_1$ are $B,G, \text{ and }D$, which are respectively mapped to nodes $f,g, \text{ and } e$, so $\tilde{\mathcal{N}}_{\graph_2}^\pi(A) = \{f,g,e\}.$

We call the neighbors of $i$'s counterpart $\mathcal{N}_{\graph_2}\big(\pi(i)\big)$.  In Fig.~\ref{fig:method}, node $A$'s counterpart is node $a$, whose neighbors are $b,g,\text{ and }d$.  Thus, $\mathcal{N}_{\graph_2}\big(\pi(A)\big) = \{b,g,d\}$.

The \textbf{matched neighborhood consistency} (MNC)~\cite{conealign} of node $i$ in $G_1$ and node $j$ in $G_2$ 
is the Jaccard similarity of the sets $\tilde{\mathcal{N}}_{\graph_2}^\pi(i)$ and $\mathcal{N}_{\graph_2}(j)$:
\begin{equation}
\small
\label{eq:jaccard}
\text{MNC}(i,j) = \frac{|\tilde{\mathcal{N}}_{\graph_2}^\pi(i) \cap \mathcal{N}_{\graph_2}(j)|}{|\tilde{\mathcal{N}}_{\graph_2}^\pi(i) \cup \mathcal{N}_{\graph_2}(j)|}.\\
\end{equation}

\subsection{Theoretical Justification of MNC.} 
\label{sec:theory}

\label{subsec:theory}

Several unsupervised network alignment algorithms attempt to enforce some notion of topological consistency in their objective functions (\S\ref{sec:related}).  We justify this intuition by showing that a specific form of topological consistency, matched neighborhood consistency or MNC, has a close relationship with alignment accuracy. 

Unsupervised network alignment is commonly evaluated on graphs that are isomorphic up to noisy or missing edges~\cite{regal,dana,magna,bigalign,final,zhang2017ineat}. When edges are removed from one graph independently with probability $p$, we show that accurate alignment entails high MNC. 

\begin{theorem}
\label{thm:mnc-noisy}
For isomorphic graphs $\graph_1 = (\vertexSet_1, \edgeSet_1)$ and $\graph_2 =(\vertexSet_2, \edgeSet_2)$, let $\pi(\cdot)$ be the isomorphism.  Let $\overline{\graph}_2 =(\vertexSet_2, \tilde{\edgeSet_2})$ be a noisy version of $\graph_2$ created by removing each edge from $\edgeSet_2$ independently with probability $p$. 
Then for any node i in $\graph_1$ and its counterpart $\pi(i)$ in $\overline{\graph}_2$,
$\mathbb{E}\big(\text{MNC}(i,\pi(i))\big) = 1 - p$.
\end{theorem}

However, this does not prove that a solution with perfect MNC will have perfect accuracy. In fact, we can construct counterexamples, such as two ``star'' graphs, each consisting of one central node connected to $n-1$ peripheral nodes (of degree one). 
Whatever the true correspondence of the peripheral nodes, aligning them to each other in any order would lead to perfect MNC. Prior network alignment work \cite{bigalign} has observed a few such special cases and in fact gives up on trying to distinguish them from the graph topology. 
We formalize this concept of \emph{structural indistinguishability}:

\begin{definition}
\label{dfn:indistinguishable}
Let $\mathcal{N}_k(u)$ be the subgraph induced by all nodes that are $k$ or fewer hops/steps away from node $u$.   Two nodes $u$ and $v$ are structurally indistinguishable if for all $k$, $\mathcal{N}_k(u)$ and $\mathcal{N}_k(v)$ are isomorphic.  
\end{definition}

Our next result proves that for isomorphic graphs, structurally indistinguishable nodes are the only possible failure case for a solution with perfect MNC.  

\begin{theorem}
\label{thm:mnc-acc}
For isomorphic graphs $\graph_1 = (\vertexSet_1, \edgeSet_1)$ and $\graph_2 =(\vertexSet_2, \edgeSet_2)$, suppose there exists $\pi(\cdot)$ that yields MNC = 1 for all nodes.  Then, if $\pi$ misaligns a node $v$ to some node $v^*$ instead of the true counterpart $v'$, it is because $v^*$ is structurally indistinguishable from $v'$.
\end{theorem}

\noindent \textbf{Proof idea}.  We give all the proofs in the supplementary \S\ref{app:proofs}.  At a high level, MNC measures the extent to which the alignment preserves edges in a node's neighborhood, and isomorphic graphs have a (not necessarily unique) perfectly edge-preserving node matching.

\noindent \textbf{Formalizing the intuitions of prior work}: 
MNC was proposed as \textit{intuition} for modeling intra-graph node proximities when performing cross-graph matching~\cite{conealign}. It was also used (not by that name) as a \textit{heuristic} in embedding-based network alignment~\cite{du2019joint}. 
Our analysis provides theoretical justification for both works.

\section{Method}
\label{sec:method}
We consider an unsupervised network alignment setting, where an initial solution, $\matchMat_0$, is provided by any network alignment method (\S\ref{sec:related}).
 We do \emph{not} take any of these initial node alignments as ground truth; on the contrary, we seek to improve their correctness by leveraging insights from our theory in \S\ref{subsec:theory}.  Thus, our problem differs from semi-supervised network alignment that is seeded with known ground-truth node correspondences~\cite{liu2016aligning,zhang2019origin}. 
Formally, we state it as:

\begin{problem}
\label{problem:na-refine}
Given a sparse initial alignment matrix $\matchMat_0$ 
between the nodes of two graphs $\graph_1$ and $\graph_2$, we seek to \emph{refine} this initial solution into a new, real-valued matrix $\matchMat$ of refined similarity scores that encodes a more accurate alignment.
\end{problem}

\subsection{\method and Connections to MNC.}

Our theoretical results 
pave a path to solving Problem~\ref{problem:na-refine} by increasing matched neighborhood consistency.  
While our results characterize ``ideal'' cases (perfect accuracy or MNC), \textit{heuristic} solutions in prior works 
have found that increasing MNC tends to increase accuracy~\cite{conealign,du2019joint}.  
Given an alignment matrix returned by a network alignment method, how can we improve its MNC in a \textit{principled} way?  We first derive a matrix-based form for MNC, which we prove in the supplementary \S\ref{app:proofs}:

\begin{theorem}
\label{thm:update}
The MNC of a binary alignment matrix $\matchMat$ can be written as a matrix $\mathbf{S}^{\text{MNC}}$ such that $\text{MNC}(i,j) = \mathbf{S}^{\text{MNC}}_{ij}$ as:

\vspace{-0.2cm}
\label{eq:mnc-matrix}
    \begin{multline}
   \mathbf{S}^{\text{MNC}} = \adj_1 \matchMat \adj_2 \; \oslash \; 
   \\
   (\adj_1\matchMat \mathbf{1}^{n_2} \otimes \mathbf{1}^{n_2} + \mathbf{1}^{n_1} \otimes \adj_2 \mathbf{1}^{n_2} - \adj_1 \matchMat \adj_2) 
   \end{multline}
where $\oslash$ is elementwise division and $\otimes$ is outer product.
\end{theorem}

We can then compute refined alignments $\matchMat'$ by multiplicative updating each node's alignment score (in $\matchMat$) with its matched neighborhood consistency: 
\begin{equation}
\label{eq:mnc-update}
\matchMat' = \matchMat \circ \mathbf{S}^{\text{MNC}}
\end{equation}
where $\circ$ denotes Hadamard product.  
Repeating over several iterations can take advantage of an improving alignment solution.  The high-level idea of our proposed refinement scheme is to \textbf{iteratively increase alignment scores for nodes that have high MNC}.

While we could just repeatedly iterate Eq.~\eqref{eq:mnc-update},  
we can introduce some mechanisms leverage important insights without increasing complexity: 

 $\bullet$ \textbf{\insightone: Prioritize high-degree nodes}. 
Higher degree nodes are easier to align~\cite{bigalign}, and so it is desirable to give them higher scores particularly in early iterations. To do so, we use only the (elementwise) numerator of Eq.~\eqref{eq:mnc-matrix}, which counts nodes' number of matched neighbors (excluding the normalizing denominator increases the score for high degree nodes).  Thus, instead of using Eq.~\eqref{eq:mnc-update}, we simplify our update rule to:
\begin{equation}
\label{eq:update}
\matchMat' = \matchMat \circ \adj_1 \matchMat \adj_2
\end{equation}
Additionally, this spares us the extra matrix operations required to compute the denominator of Eq.~\eqref{eq:mnc-matrix}.  

 $\bullet$ \textbf{\insighttwo: Do not overly rely on the initial solution}. 
At every iteration, we add a small $\epsilon$ to every element of $\matchMat$.  This gives each pair of nodes a token match score whether or not the initial alignment algorithm identified them as matches, which gives us a chance to correct the initial solution's false negatives.  

 $\bullet$ \textbf{\insightthree: Allow convergence.}
Finally, to keep the scale of the values of $\matchMat$ from exploding, we 
we row-normalize $\matchMat$ followed by column-normalizing it at every iteration. 
Previous iterative graph matching methods such as graduated assignment~\cite{gold1996graduated} require full normalization per iteration using the Sinkhorn algorithm~\cite{sinkhorn1964relationship} to produce a doubly stochastic matrix.  In contrast, 
with \method a single round of normalization suffices, avoiding large computational expense (\S\ref{sec:experiments}). 

Putting it all together, we give the pseudocode of our method \method in Algorithm~\ref{alg:refinement}.  
\method is powerful yet conceptually simple and straightforward to implement.  It requires only a few lines of code, with each line implementing an important insight. 

\subsection{Connections to Other Graph Problems.}
In the supplementary \S\ref{app:connections}, we further justify \method conceptually by comparing and contrasting it to several other graph techniques. We find that \method's update can be viewed as a more flexible version of the seed-and-extend node matching heuristic~\cite{graal,patro2012global}, and that it performs a graph filtering operation akin to a graph neural network~\cite{sgc}. In particular, a similar neighborhood aggregation operation is used by graph neural tangent kernels~\cite{gntk} for graph-level comparison.  This further highlights the connection between the individual node similarities found in graph matching~\cite{final,regal} and aggregated node similarities in graph kernels~\cite{final,rgm}.

{\footnotesize
\begin{algorithm}[t!]
\caption{\method($\adj_1, \adj_2, \matchMat_0, K, \epsilon$)} 
\label{alg:refinement}
\begin{algorithmic}[1]
\State \textbf{Input:}  adjacency matrices $\adj_1, \adj_2$, initial alignment matrix $\matchMat_0$, number of iterations $K$, token match score $\epsilon$
\For{$k = 1 \to K$}  \Comment{Refinement iterations} 
    \State $\matchMat_k = \matchMat_{k-1} \circ \adj_1 \matchMat_{k - 1} \adj_2$ \Comment{MNC update} 
    \State $\matchMat_k = \matchMat_k + \epsilon$ \Comment{Add token match scores} 
    \State $\matchMat_k = \text{Normalize}(\matchMat_k)$ \Comment{By row then column}
     
\EndFor
\State \Return{$\matchMat_K$} 
\end{algorithmic}
\end{algorithm}
}

\subsection{Optimizations: Sparse \method.}
\label{subsec:sparse}

The dense matrix updates lead to quadratic computational complexity in the number of nodes. 
To scale \method to large graphs, we sparsify it by updating only a small number of alignment scores for each node.  Intuitively, we forgo updating scores of likely non-aligned node pairs (these updates would be small anyway). 
Concretely, sparse \method replaces Line 3 of Algorithm~\ref{alg:refinement} with 
$$\matchMat_{k|\updateMatch_k} = \matchMat_{k-1|\updateMatch_k} \circ \updateMatch_k$$ 
where the update matrix $\updateMatch_k = \text{top-}\alpha(\adj_1 \matchMat \adj_2)$ is a sparse version of $\adj_1 \matchMat \adj_2$ containing only the largest $\alpha$ entries per row.    $\matchMat_{k|\updateMatch_k}$ selects the elements of $\matchMat_k$ (pairs of nodes) corresponding to nonzero elements in $\updateMatch_k$. These are the only elements on which we perform an MNC-based update, and  the only ones to receive a token match score (Line 4): $\matchMat_{k|\updateMatch_k} = \matchMat_{k|\updateMatch_k} + \epsilon$.   

As the elements to update are selected by the size of their update scores, which are computed using the previous solution $\matchMat$, sparse \method relies somewhat more strongly on the initial solution.  However, we still comply with \textbf{\insighttwo} by updating $\alpha > 1$ possible alignments for each node.  This has sub-quadratic time and space requirements (cf. supplemental \S\ref{app:complexity}) and accuracy comparable to dense updates (\S~\ref{sec:exp-scalability}).

\subsection{Assumption and Limitations.}
\label{sec:limitations}
Network alignment typically relies on some kind of topological consistency assumption (\S\ref{sec:related}).  Likewise, \method assumes that improving alignment MNC will improve its quality.  Our theory (\S\ref{sec:theory}) shows this assumption is well-founded for the quasi-isomorphic graphs studied by many prior works~\cite{regal,dana,magna,bigalign,final,zhang2017ineat}, and our experiments (\S\ref{sec:experiments}) support \method's efficacy in even more use cases.  
However, all assumptions have limitations, so we discuss two limitations of ours.  
First, as Theorem~\ref{thm:mnc-acc} notes, misaligned nodes can have high MNC. In practice, we find that this is comparatively rare, and we shed insight into when it may occur (\S\ref{sec:exp-insightone}). 
Second, the correct alignment may have low MNC, as is the case in some datasets consisting of multiple social networks sharing a limited number of common users~\cite{final,liu2016aligning}.  Recent work characterizes these datasets as having low \emph{structural credibility}: topological consistency widely breaks down, and supervision and/or rich attribute information is generally needed for successful network alignment~\cite{wang2020credible}. 
Future work could extend \method to use such information.

\section{Experiments}
\label{sec:experiments}
In this section, we first demonstrate \method's ability to improve diverse unsupervised network alignment methods in a variety of challenging scenarios at reasonable computational cost. We next perform a deep study of \method that verifies its various design insights. 

\subsection{Experimental Setup}
\label{subsec:setup}

\subsubsection{Data.} We choose network datasets from a variety of domains (e.g. biological, social) (Tab.~\ref{tab:datasets}). We consider two scenarios for network alignment with \textbf{ground truth node correspondence}:
{\bf (1) \textit{Simulated noise.}} 
For each network with adjacency matrix $\adj$, we create a permuted copy $\tilde{\adj} = \mathbf{P} \adj \mathbf{P}^\top$, where $\mathbf{P}$ is a random permutation matrix, and add noise by removing each edge with probability $p \in [0.05, 0.10, 0.15, 0.20, 0.25]$.
The task is to align each network to its \emph{noisy} permuted copy ${\tilde{\adj}}^{(p)}$, with the ground truth alignments being given by $\mathbf{P}$.      
{\bf (2) \textit{Real noise.}} Our PPI-Y dataset is commonly studied in biological network alignment~\cite{magna}.  
The task is to align the largest connected component of the yeast (\emph{S.cerevisiae}) PPI network to its copies augmented with 5, 10, 15, 20, and 25 percent additional low-confidence PPIs (added in order of their confidence)~\cite{magna}.

We also show (\S\ref{sec:exp-performance-real}) that \method helps find more meaningful alignments even for networks defined on differing underlying nodes, by aligning graphs with \textbf{no ground truth node correspondence}: the PPI networks of \textbf{(3) \textit{separate species}} (SC \& DM).  We align a yeast network (\emph{S.cerevisiae}, a larger network than PPI-Y) to a fruit fly network (\emph{D.melanogaster}).

\begin{table}[t!]
\centering
\caption{Description of the datasets used.} 
\label{tab:datasets}
\vspace{-0.2cm}
{\footnotesize
\begin{tabular}{l r@{\hspace{0pt}}  r@{\hspace{1pt}} @{\hspace{1pt}}r l}
\toprule
   \textbf{Name} & \textbf{Nodes} & \textbf{Edges} &  \textbf{Description}  \\
\midrule
     Arenas Email \cite{koblenz} &  1\,133      & 5\,451    & communication \\
     
     Hamsterster \cite{koblenz} & 2\,426 & 16\,613 & social \\
     
     PPI-H \cite{ppi} & 3\,890     & 76\,584    &  PPI (human) \\
     
     Facebook \cite{snapnets} & 4\,039 & 88\,234 & social \\ 
 PPI-Y~\cite{magna} & 1\,004 & 8\,323 & PPI (yeast) \\ 
     LiveMocha \cite{koblenz} & 104\,103  & ~2\,193\,083 & social \\
     
     \midrule
     
     \multirow{2}{*}{SC \& DM \cite{isorank}} & 5\,926 & 88\,779 & PPI (yeast) \\
     
     & 1\,124 & 9\,763 & PPI (fruit fly) \\

\bottomrule
\end{tabular}
}
\vspace{-0.2cm}
\end{table}

\subsubsection{Metrics.} For graphs with ground-truth node correspondence, our main measure of alignment success is \textbf{accuracy}: the proportion of correctly aligned nodes.
While this is the primary goal of network alignment, we also give \textbf{matched neighborhood consistency} (MNC, Eq.~\eqref{eq:jaccard}) as a secondary metric (e.g. in Fig.~\ref{fig:deg-before}-b), in light of its importance in our analysis (\S\ref{sec:theory}). 

When the networks have no ground-truth node correspondence, accuracy is not defined, so we assess two properties of the conserved network $\overline{\adj} = \matchMat^\top \adj_1 \matchMat \circ \adj_2$ (the overlap of the aligned adjacency matrices):

\begin{itemize}
    \item \textbf{Normalized Overlap (N-OV)}: the percentage of conserved edges: $100*\frac{\text{nnz}(\overline{\adj})}{\max(\text{nnz}(\adj_1), \text{nnz}(\adj_2))}$
    \item \textbf{Largest Conserved Connected Component (LCCC)}: the number of edges in the largest connected component of $\overline{\adj}$. 
\end{itemize}

These respectively measure the alignment solution's ability to conserve edges and large substructures, which in biological networks may correspond to functions that are shared across species~\cite{isorank}.  Thus, larger values may indicate a more biologically interesting alignment.   

\subsubsection{Base Network Alignment Methods.} 
To illustrate \method's flexibility, we apply it to a set of base network alignment methods using a diverse range of techniques (belief propagation, spectral methods, genetic algorithms, and node embeddings):
\textbf{(1)~NetAlign}~\cite{netalign}, \textbf{(2) FINAL}~\cite{final}, \textbf{(3) REGAL}~\cite{regal}, \textbf{(4) CONE-Align}~\cite{conealign},
and \textbf{(5) MAGNA}~\cite{magna}.  

  \begin{figure*}[t!]
	\centering
    \subfloat{
      \includegraphics[width=.82\textwidth]{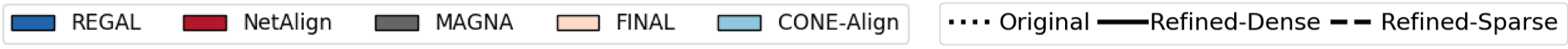}
    }
	\vspace{-0.35cm}

    \setcounter{subfigure}{0}
    \subfloat[Arenas\label{fig:acc-arenas}]{
      \includegraphics[width=0.18\textwidth, trim=0 0 1.5cm 1.5cm, clip]{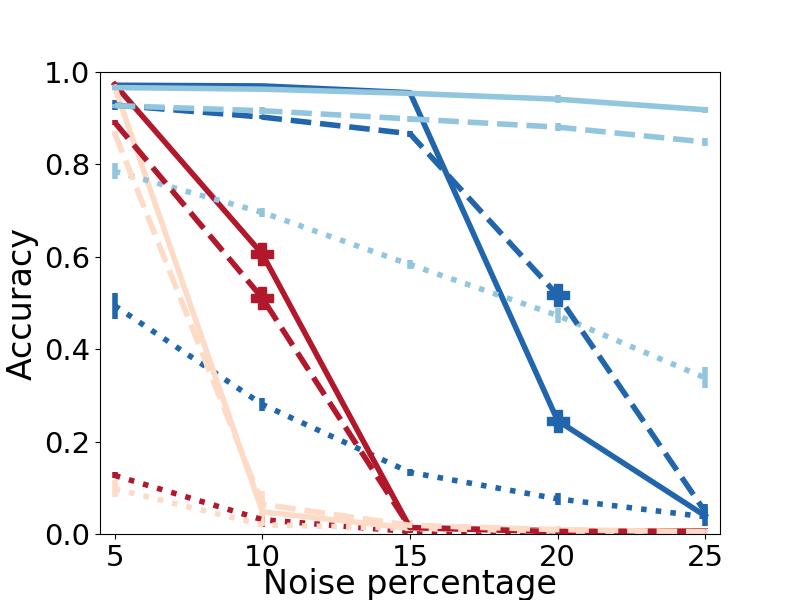}
    }
    ~
    \subfloat[Hamsterster\label{fig:acc-hamster}]{
      \includegraphics[width=0.18\textwidth, trim=0 0 1.5cm 1.5cm, clip]{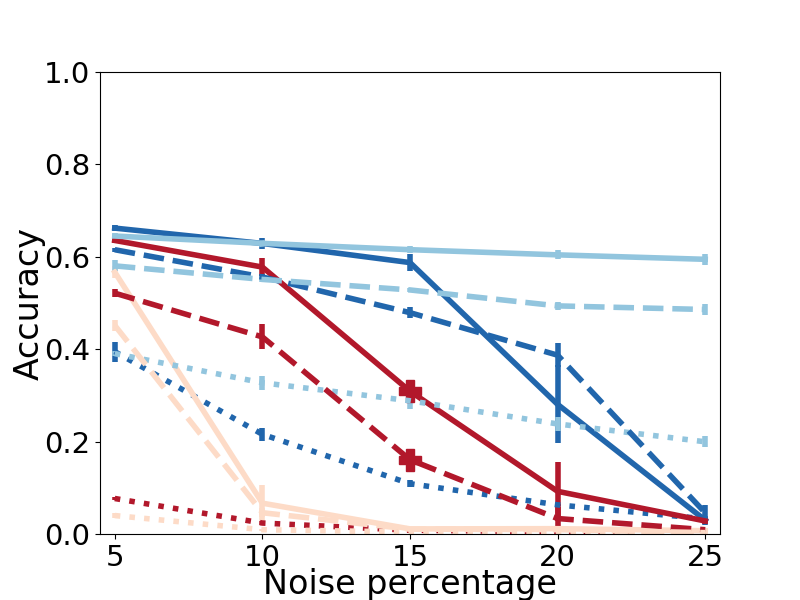}
    }
    ~
    \subfloat[PPI-H\label{fig:acc-ppi}]{
      \includegraphics[width=0.18\textwidth, trim=0 0 1.5cm 1.5cm, clip]{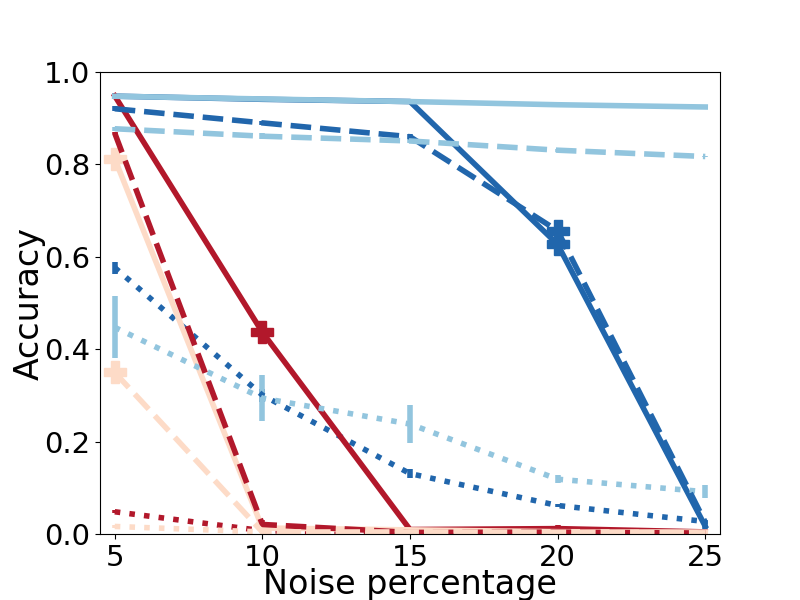}
    }
    ~
    \subfloat[Facebook
    \label{fig:acc-fb}]{\includegraphics[width=0.18\textwidth, trim=0 0 1.5cm 1.5cm, clip]{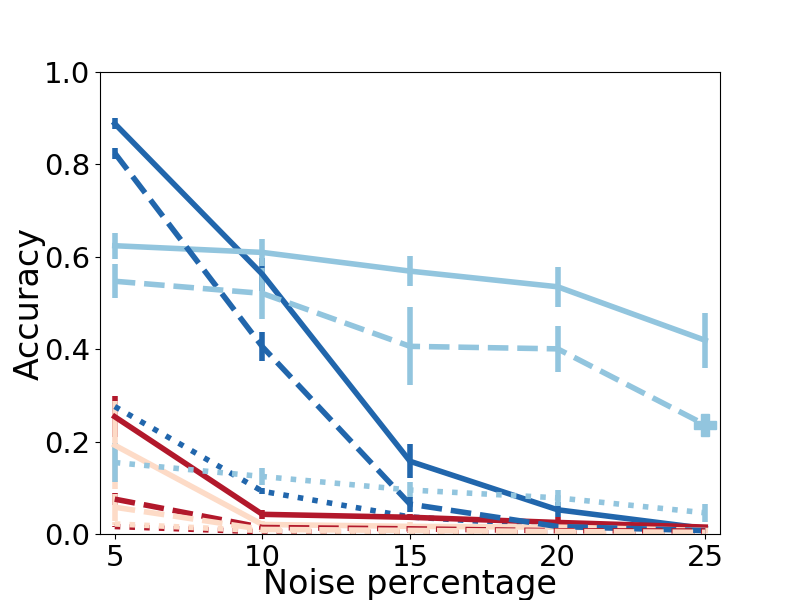}
    }
    ~
    \subfloat[PPI-Y
    \label{fig:acc-magna}]{\includegraphics[width=0.18\textwidth, trim=0 0 1.5cm 1.5cm, clip]{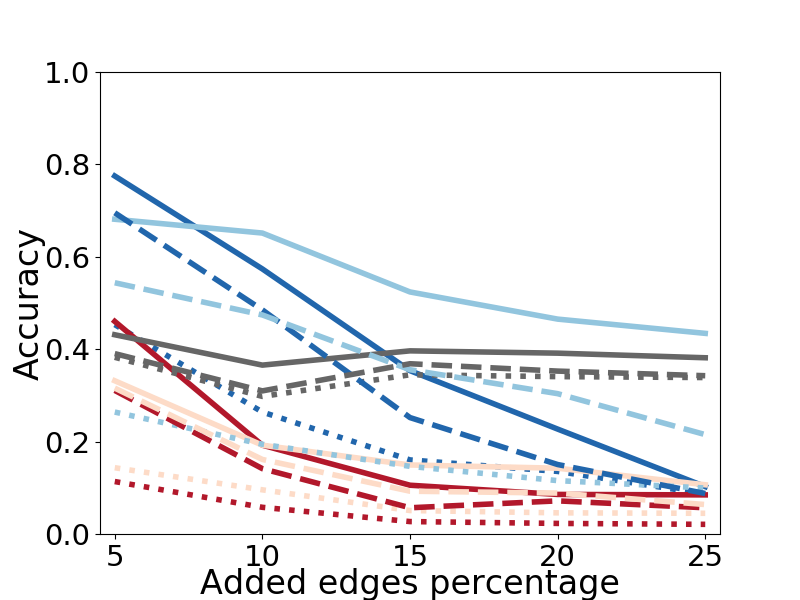}
    }
	\caption{Alignment accuracy vs graphs' topological difference.  
	Both sparse and dense refinement with \method improve all base methods' alignment accuracy and robustness, often quite dramatically. 
	(We run MAGNA only on PPI-Y, the dataset with real noise, due to its high runtime---cf.\ Fig.~\ref{fig:simulated-run}\protect\subref{fig:run-magna}.)}
    \label{fig:simulated-acc} 
\end{figure*}

\begin{figure*}[t!]
	\centering
	\vspace{-0.15cm}
    \subfloat{
      \includegraphics[width=.82\textwidth]{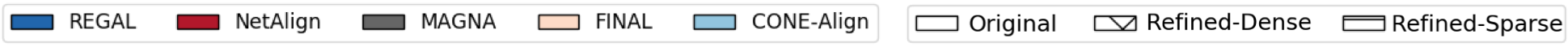}
    }
	\vspace{-0.35cm}
	
    \setcounter{subfigure}{0}
    \subfloat[Arenas\label{fig:run-arenas}]{
      \includegraphics[width=0.19\textwidth, trim=0 0.8cm 1.5cm 1.5cm, clip]{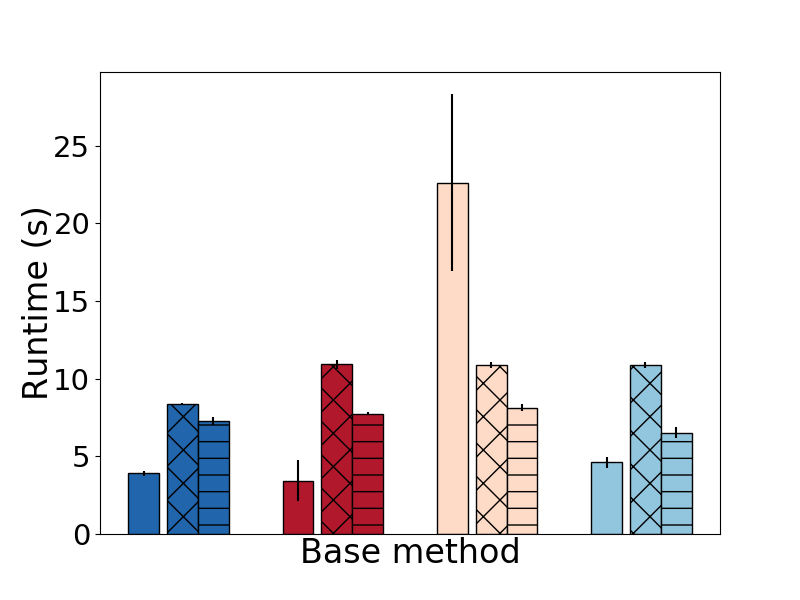}
    }
    \subfloat[Hamsterster\label{fig:run-hamster}]{
      \includegraphics[width=0.19\textwidth, trim=0 0.8cm 1.5cm 1.5cm, clip]{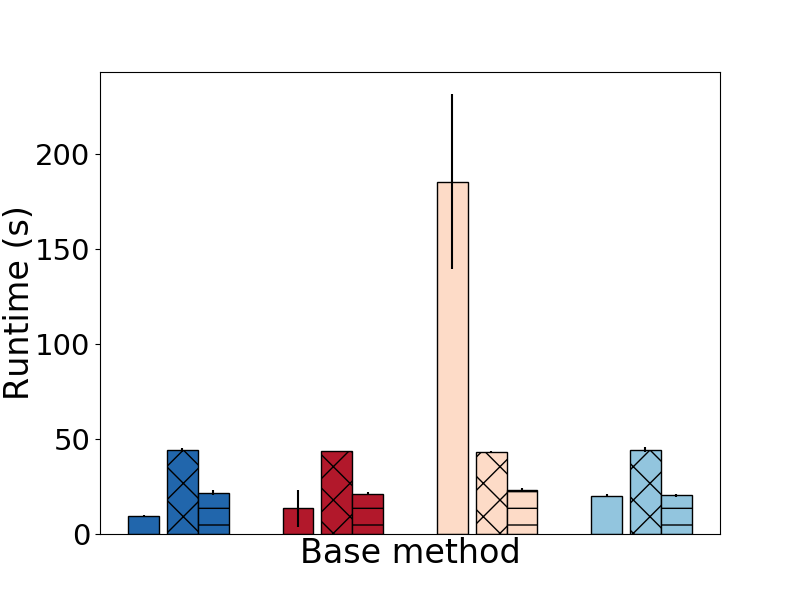}
    }
    \subfloat[PPI-H\label{fig:acc-ppi}]{
      \includegraphics[width=0.19\textwidth, trim=0 0.8cm 1.5cm 1.5cm, clip]{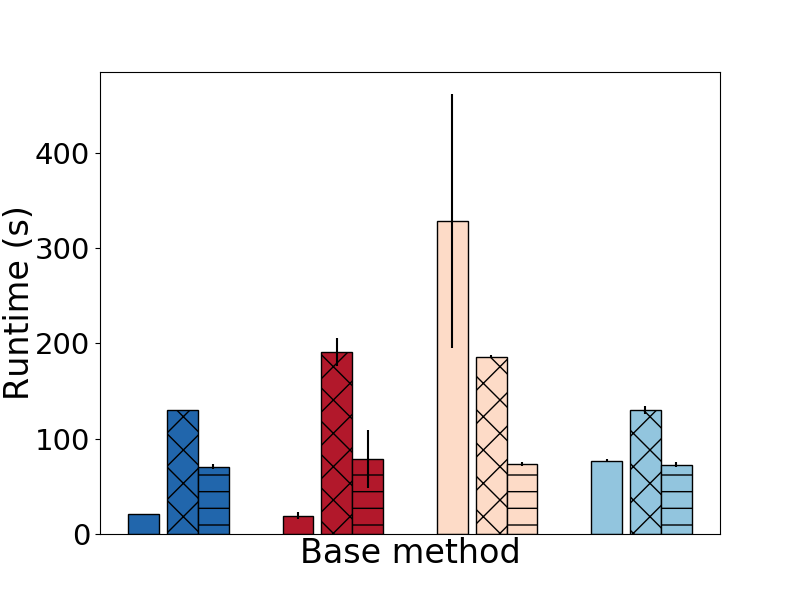}
    }
    \subfloat[Facebook
    \label{fig:run-fb}]{\includegraphics[width=0.19\textwidth, trim=0 0.8cm 1.5cm 1.5cm, clip]{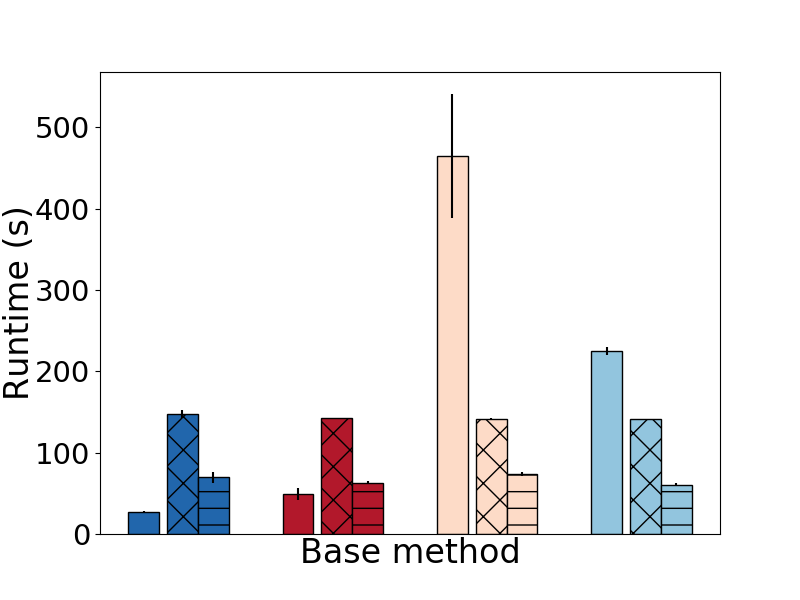}
    }
    \subfloat[PPI-Y
    \label{fig:run-magna}]{\includegraphics[width=0.19\textwidth, trim=0 0.8cm 1.5cm 1.5cm, clip]{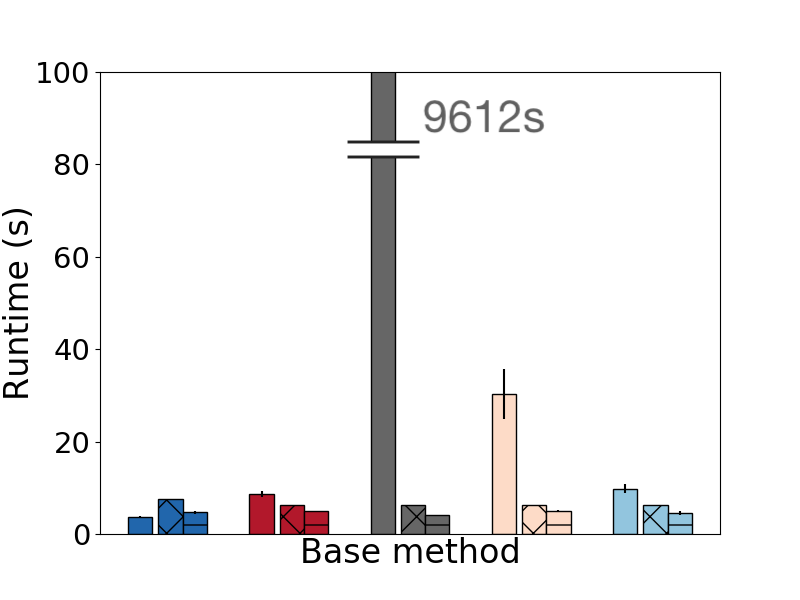}
    }
    \vspace{-0.2cm}
	\caption{Runtime for different refinement variations.  Sparse refinement is appreciably faster than dense refinement.  Both refinements offer a modest computational overhead, often on par with or faster than the time taken to perform the original network alignment.}
    \label{fig:simulated-run}
	\vspace{-0.15cm}    
\end{figure*}

\noindent $\bullet$ \textit{Base Method Settings.}
We configure all base methods following the literature (cf. the supplementary \S\ref{app:params}). 

\noindent $\bullet$ \textit{\method Settings.}  By default, we use $K=100$ refinement
iterations (cf.\ supplementary \S\ref{app:convergence}) and token match score
$\epsilon=\frac{1}{ 10^{\overline{p}} }$ 
where $\overline{p} = \min\{p \in \mathbb{N}: 10^p > n\}$, 
so that a node's token match scores will sum to less than 1.   
For sparse refinement, we update $\alpha=10$ entries per node.  We justify these choices by sensitivity analysis.

\vspace{-0.2cm}
\subsection{Analysis of Alignment Performance}
\subsubsection{Finding Known Correspondences.}
\label{sec:exp-performance-syn}
In Fig.~\ref{fig:simulated-acc} we plot the accuracy of all methods with (solid/dashed lines) and without (dotted lines) refinement.  For simulated experiments (\ref{fig:acc-arenas}-\ref{fig:acc-fb}) we report average and standard deviation of accuracy over five trials.  

\vspace{0.1cm}
\noindent \textit{Results.} We see dramatic improvements in alignment accuracy for \emph{all} methods with refinement.   A striking example is NetAlign on the PPI-H dataset with 5\% noise.  Initially finding just 4\% of alignments, with \method it achieves 94\% accuracy--going from a nearly completely incorrect to nearly completely correct.  Similarly, CONE-Align achieves well over 90\% accuracy on Arenas and PPI-H with refinement \emph{and} sees only a slight drop even at the highest noise levels.

\begin{observation}
\method greatly improves the accuracy of diverse network alignment methods across datasets.  
\end{observation}

We are also able to align networks that are much noisier than previous works had considered: our \emph{lowest} noise level is 5\%, the \emph{highest} noise level in~\cite{regal}. \method's benefit is appreciable at up to 10\% noise on most datasets for FINAL, 15\% noise for NetAlign, 20\% for REGAL, and the full 25\% noise for CONE-Align. 

 \begin{observation}
 Refinement can make network alignment methods considerably more robust to noise.
 \end{observation}

Fig.~\ref{fig:simulated-acc} and the supplementary Tab.~\ref{tab:results} (\S\ref{app:table}) show that network alignment methods vary in how much their accuracy increases from \method, and also the maximum noise level at which they appreciably improve. This variation shows that \method is not ignoring the initial solution.   Precisely characterizing the ``refinability'' of initial solutions is an interesting question for future work.  Here, our goal is not to rank ``better'' or ``worse'' existing methods; instead, comparing all methods to \emph{their own} unrefined solutions shows the value of \method.

\begin{observation}
Different methods benefit differently from refinement, but all benefit considerably.
\end{observation}

Compared to dense refinement, sparse refinement is slightly less accurate overall.  However, the gap between the two variants decreases as the noise level increases, and in some cases (e.g., REGAL on Arenas and Hamsterster at high noise levels) sparse refinement even performs better, possibly indicating a regularizing effect of forgoing updates of low-confidence node pairs.  Sparse refinement is faster than dense refinement, as Fig.~\ref{fig:simulated-run} shows, but both offer a reasonable overhead compared to the initial alignment time.  In general, they are modestly slower than NetAlign and REGAL, two famously scalable methods, on par with CONE-Align, faster than FINAL, and much faster than MAGNA.\footnote{On account of MAGNA's high runtime, we only run it on PPI-Y (where it takes 9612 seconds on average.)}

\begin{observation}
\method has a manageable computational overhead, particularly with sparse updates.
\end{observation} 

\subsubsection{Conserving Meaningful Structures.}
\label{sec:exp-performance-real}

We now use \method to help REGAL, NetAlign, and FINAL~\footnote{CONE-Align's subspace alignment operates on embeddings for the same number of nodes, which these graphs do not have.} align the SC-DM dataset consisting of the PPI networks of different species: graphs whose nodes do not necessarily have an underlying correspondence.

\begin{figure}[t!]
	\centering
	\vspace{-0.2cm}
    \subfloat{
      \includegraphics[width=.2\textwidth]{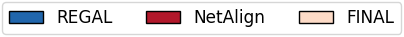}
    }
	~
	\subfloat{
      \includegraphics[width=.27\textwidth]{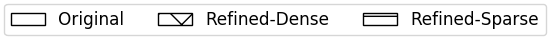}
    }
	\vspace{-0.35cm}
	
    \setcounter{subfigure}{0}
    \subfloat[Normalized Overlap\label{fig:run-arenas}]{
      \includegraphics[width=0.19\textwidth, trim=0 0.8cm 1.5cm 1.5cm, clip]{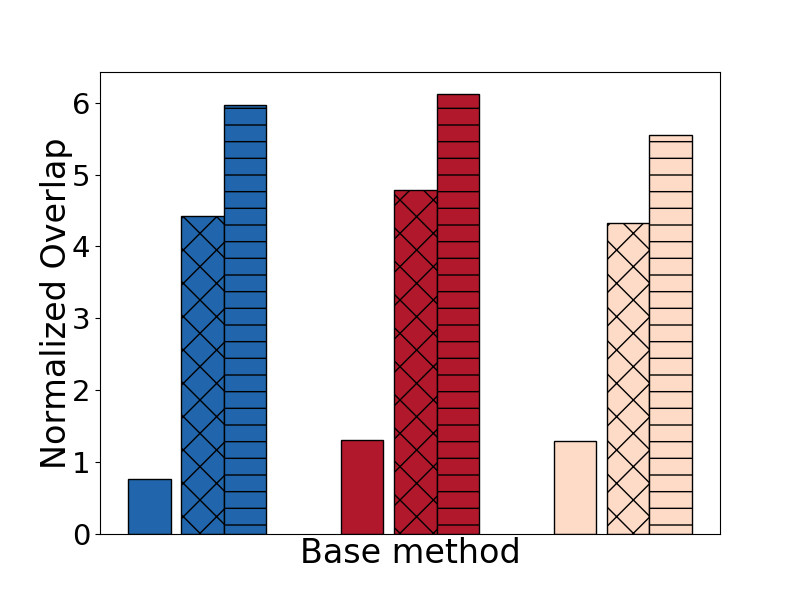}
    }
    \subfloat[Size of LCCC\label{fig:run-hamster}]{
      \includegraphics[width=0.19\textwidth, trim=0 0.8cm 1.5cm 1.5cm, clip]{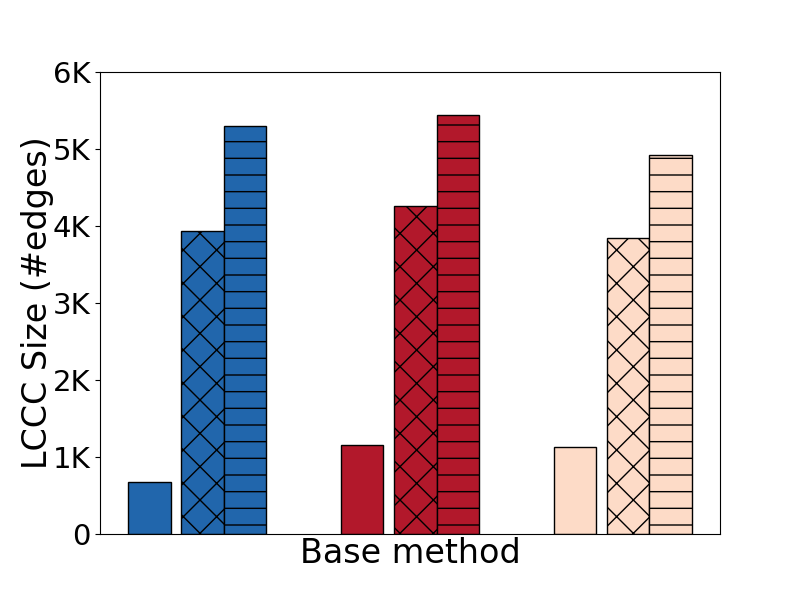}
    }
    \vspace{-0.15cm}
	\caption{Alignment of PPI networks of different species with no ground truth node alignment. \method helps base methods find potentially more biologically interesting solutions by multiple metrics.}
    \label{fig:nonisomorphic}
	\vspace{-0.15cm}    
\end{figure}

\noindent \textit{Results.}  Fig.~\ref{fig:nonisomorphic} shows that using \method, both measures of structural conservation increase significantly for all methods, indicating more successful alignments.  In fact, these results are on par with ones obtained using amino acid sequence information in addition to graph topology~\cite{isorank}. 
Interestingly, sparse refinement brings even greater improvements than dense. 

\begin{observation}
\method is useful even for graphs with no underlying node correspondence.
\end{observation}

\subsection{Drilldown: Network Alignment Insights.}
\label{sec:sensitivity}
In this section, we perform a drilldown of (dense) \method that gives a deeper understanding into its specific design choices in light of the three insights we laid out in \S\ref{sec:method}.

\begin{figure*}[t!]
    \vspace{-0.75cm}
	\centering
    \subfloat[Distribution of MNC and accuracy by node degree \textit{before} refinement: NetAlign, Arenas 5\% noise. \label{fig:deg-before}]{
      \includegraphics[width=0.21\textwidth]{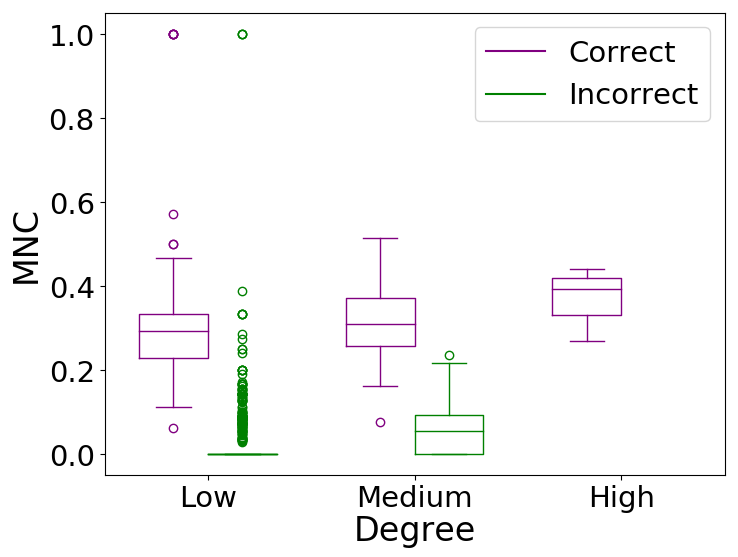}
    }
    \hfill
    \subfloat[Distribution of MNC and accuracy by node degree \textit{after} refinement: NetAlign, Arenas 5\% noise.. \label{fig:deg-after}]{
      \includegraphics[width=0.21\textwidth]{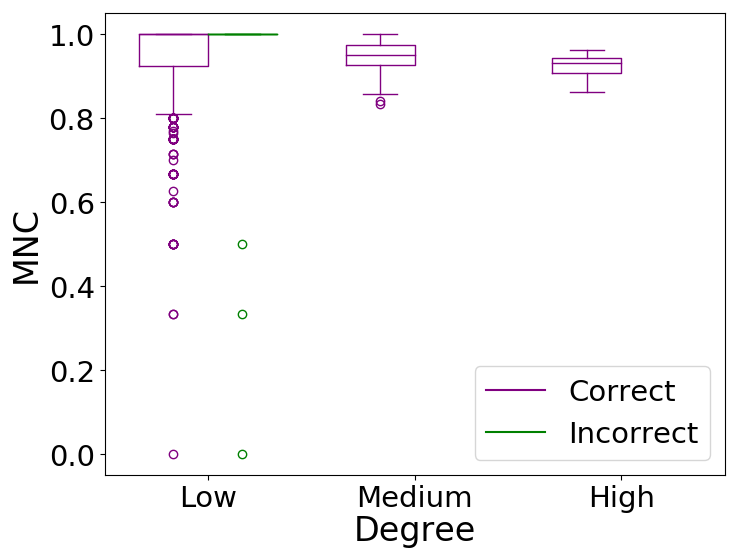}
    }
    \hfill 
    \subfloat[Accuracy with varying token match score $\epsilon$: Arenas, 5\% noise. The methods almost overlap for $\epsilon > 0$. \label{fig:sensitivity:eps}]{
      \includegraphics[width=0.215\textwidth]{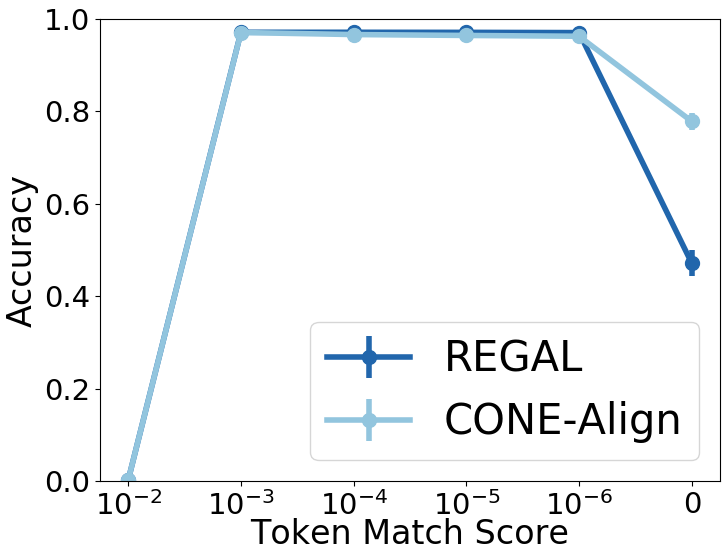} 
    }
    \hfill
    \subfloat[Accuracy/runtime per iteration with limited vs full Sinkhorn normalization: NetAlign, Hamsterster 5\% noise. \label{fig:sensitivity:normalization}]{
      \includegraphics[width=0.24\textwidth]{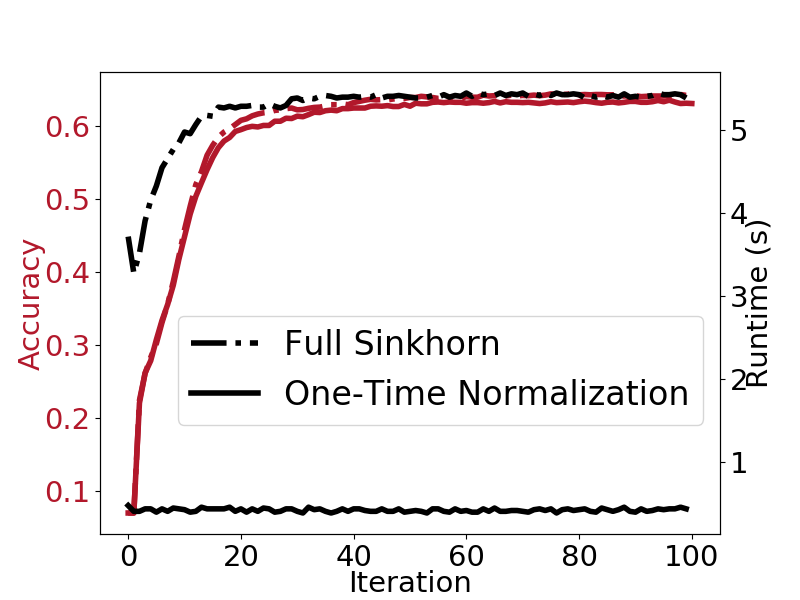}
    }
	\vspace{-0.15cm}
	\caption{Drilldown of \method in terms of our three insights that inspired its design.  \protect\subref{fig:deg-before}-\protect\subref{fig:deg-after} For \insightone, before and after alignment, high degree nodes are more likely to have high MNC and be correctly aligned.   \protect\subref{fig:sensitivity:eps} For \insighttwo, our token match score admits a wide range of values that yield good refinement performance.  \protect\subref{fig:sensitivity:normalization} For \insightthree, our proposed normalization effectively avoids the computational expense of full Sinkhorn normalization.}
    \label{fig:lessons}
    \vspace{-0.25cm}
\end{figure*}

\subsubsection{Aligning High Degree Nodes \emph{\bf (\insightone)}.}
\label{sec:exp-insightone}
In Figs.~\ref{fig:deg-before}-b, 
we analyze the claim that high-degree nodes are easier to align, which inspires \method's design, for NetAlign on the Arenas dataset for brevity. 
We split the nodes into three groups by degree: $[0, \frac{\degmax}{3}), [\frac{\degmax}{3}, \frac{ 2\degmax}{3}), [\frac{ 2\degmax}{3},  \degmax]$ ($\degmax$ is the maximum node degree) and plot the distribution of MNC in each group among correct and incorrectly aligned nodes.

\vspace{0.1cm}
\noindent \textit{Results.} 
 High-degree nodes are rarely misaligned even before refinement, and the few nodes that are still misaligned afterward have low degrees.  These often still have a high MNC, probably because it is easier for smaller node neighborhoods to be (nearly) isomorphic, resulting in (near) structural indistinguishability (the challenge indicated by Theorem~\ref{thm:mnc-acc}.)  
 
  \begin{observation}
 It is easier to align high-degree nodes, verifying \insightone that motivates \method's update formulation to encourage early alignment of high-degree nodes.  
 \end{observation}

\subsubsection{Token Match Score \emph{\bf (\insighttwo)}.}
We now study the effects of the token match score $\epsilon$, used to overcome some limitations of an erroneous initial solution (our second network alignment insight in \S\ref{sec:method}).  We use the Arenas dataset with 5\% noise averaged over five trials (we observe similar trends on other datasets), varying $\epsilon$ from $10^{-2}$ to $10^{-6}$ and $\epsilon = 0$ (no token match score).

\vspace{0.1cm}
\noindent \textit{Results.} In Fig.~\ref{fig:sensitivity:eps}, we see that the performance can dramatically drop with too large $\epsilon$, where the token match scores overwhelm the alignment information.  
We need a positive token match score for the multiplicative update rule to work: with $\epsilon = 0$, \method fails to discover new alignments and only achieves the initial solutions' accuracy (cf. Fig.~\ref{fig:acc-arenas}). However, \method works with a wide range of small positive choices of $\epsilon$, including $\epsilon = 10^{-4}$ as recommended by our criterion in \S~\ref{subsec:setup}. 

 \begin{observation}
\method is robust to the token match score, as long as it is not so large that it would drown out the actual node similarity scores.   
 \end{observation}

\subsubsection{Alignment Matrix Normalization \emph{\bf (\insightthree)}.}
\method normalizes the rows and columns of the alignment matrix, per our third insight in \S\ref{sec:method}.  Sinkhorn's algorithm iteratively repeats this process, converging to a doubly stochastic matrix~\cite{sinkhorn1964relationship}.  Methods such as graduated assignment~\cite{gold1996graduated} nest iterative Sinkhorn normalization at every iteration of optimization.  We refine NetAlign's solution on the Hamsterster dataset with 5\% noise, both as proposed and using Sinkhorn's algorithm (up to 1000 iterations or a tolerance of $10^{-2}$).

\vspace{0.1cm}
\noindent \textit{Results.}   We see in Fig.~\ref{fig:sensitivity:normalization} that Sinkhorn's algorithm takes longer to converge (particularly as refinement continues) and dominates the running time.  Meanwhile, our proposed normalization yields virtually the same accuracy, while keeping computation time low ($\ll 1$ second per iteration) and relatively fixed. 

\begin{observation}
\method is advantageous by being able to eschew the full Sinkhorn procedure.   
 \end{observation}

\begin{figure}[t!]
    \vspace{-0.55cm}
    \subfloat[Accuracy/runtime vs sparsity level $\alpha$: CONE-Align, Facebook 5\% noise.
    \label{fig:sensitivity:sparse-accrun}]{
      \includegraphics[width=0.22\textwidth]{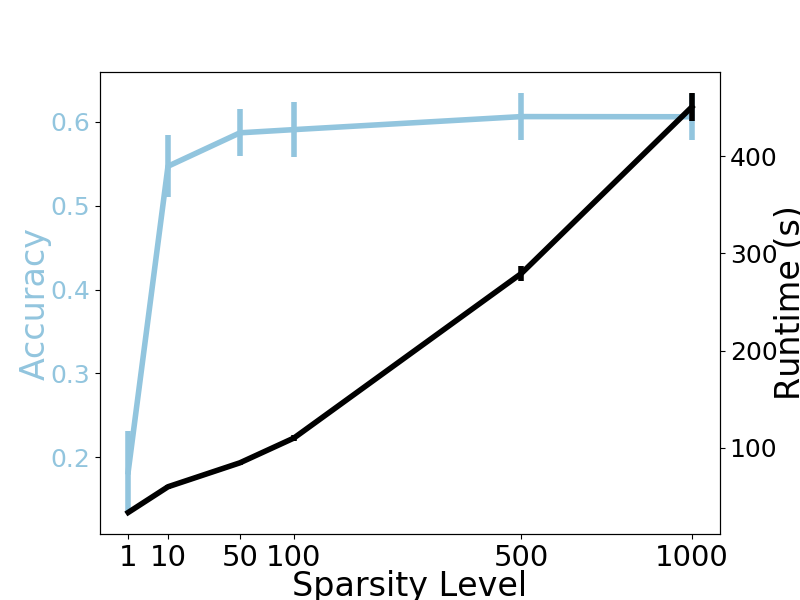}
    }
    \hfill
    \subfloat[Top-$\numTop$ accuracy on a time budget: REGAL, LiveMocha 5\% noise. \label{fig:sensitivity:topalpha}]{\includegraphics[width=0.22\textwidth]{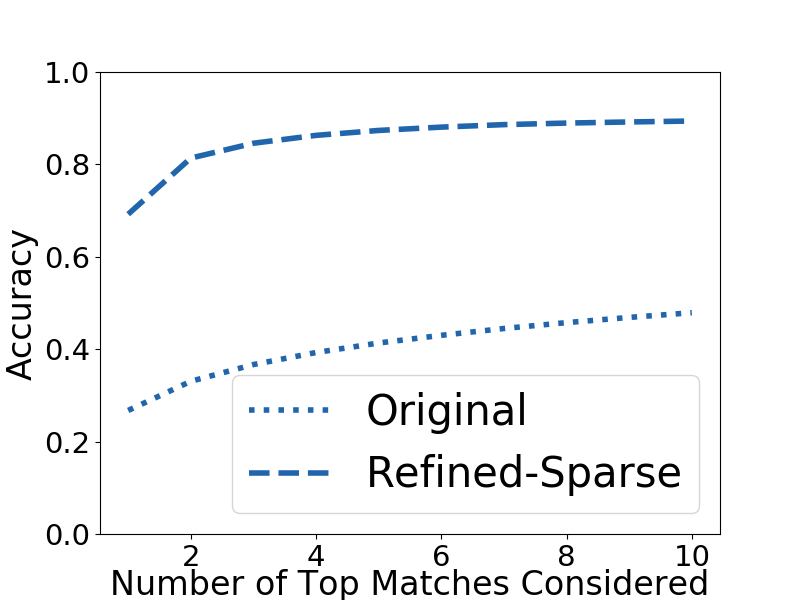}
    }
    \caption{Sparse \method.  
    \protect\subref{fig:sensitivity:sparse-accrun} Varying the update sparsity allows us to trade off accuracy and runtime.
    \protect\subref{fig:sensitivity:topalpha} On extremely large graphs, \method uncovers additional alignment (and near-alignments).}
    \label{fig:sensitivity}
    \vspace{-0.1cm}
\end{figure}

\subsection{Sparse Updates and Scalability}
\label{sec:exp-scalability}

\subsubsection{Sparsity Level of Refinement.}
By varying $\alpha$, the number of updates per node, we can interpolate between the sparse and dense versions of \method.  We study this on Facebook with 5\% noise.

\vspace{0.1cm}
\noindent \textit{Results.}  
Updating just one alignment per node leads to poor performance, as per \textbf{\insighttwo}: we should not blindly trust the initial solution by using only its top choice.  However, the initial solution provides enough information that we can use only the top few choices, with marginal accuracy returns compared to the extra runtime required by using more top choices.  Thus, sparse \method offers a favorable balance of accuracy and speed.  

\subsubsection{``Soft'' Alignments, Large Graphs.}
To utilize the scalability of sparse \method, we use it on a large dataset: LiveMocha, which has over 100K nodes and a \emph{million}  edges. 
We simulate an alignment scenario with 5\% noise.  We only use REGAL, the most scalable base alignment method we consider, together  with sparse refinement (dense refinement runs out of memory).  We consider a \emph{budgeted computation} scenario, running \method for as long as REGAL takes (2600s). Meanwhile, we explore the top-$\numTop$ accuracy: the proportion of correct alignments in the top $\numTop$ choices per node~\cite{regal} ranked by the real-valued entries of $\matchMat$.

\vspace{0.1cm}
\noindent \textit{Results.}  In Fig.~\ref{fig:sensitivity:topalpha}, we see that \method scales to large graphs and more than doubles the accuracy of REGAL in the same amount of time it took REGAL to obtain its initial solution.  The top-$\numTop$ scores also increase as we consider more top matches per node (up to the sparsity parameter 10), which may be useful for some applications~\cite{regal}.

\begin{observation}
Sparse refinement offers a favorable efficiency tradeoff and scales to large graphs.  
 \end{observation}

\section{Conclusion}
\label{sec:conclusion}
We have proposed \method, a powerful technique for refining existing network alignment methods that greatly improves accuracy and robustness.  \method is simple to implement and apply \emph{post hoc} to a variety of methods.  Its compact formulation encodes  several insights for network alignment, supported by our extensive theoretical and empirical analysis, and is highly scalable with sparse computation.  We hope that all these positive qualities will make it attractive to practitioners.

\section*{Acknowledgements}
{\small
This work is supported by
NSF Grant No.\ IIS 1845491, Army Young Investigator Award No.\ W9-11NF1810397, 
and Adobe, Amazon, Facebook, and Google faculty awards.
}

\balance
\bibliographystyle{plain}
\bibliography{BIB/abbrev,BIB/bibliography}

\begin{thebibliography}{10}

\bibitem{aflalo2015convex}
Yonathan Aflalo, Alexander Bronstein, and Ron Kimmel.
\newblock On convex relaxation of graph isomorphism.
\newblock {\em PNAS}, 2015.

\bibitem{netalign}
Mohsen Bayati, Margot Gerritsen, David~F Gleich, Amin Saberi, and Ying Wang.
\newblock Algorithms for large, sparse network alignment problems.
\newblock In {\em ICDM}, 2009.

\bibitem{ppi}
Bobby-Joe Breitkreutz, Chris Stark, Teresa Reguly, Lorrie Boucher, Ashton
  Breitkreutz, Michael Livstone, Rose Oughtred, Daniel~H Lackner, J{\"u}rg
  B{\"a}hler, Valerie Wood, et~al.
\newblock The biogrid interaction database: 2008 update.
\newblock {\em Nucleic acids research}, 2008.

\bibitem{conealign}
Xiyuan Chen, Mark Heimann, Fatemeh Vahedian, and Danai Koutra.
\newblock Cone-align: Consistent network alignment with proximity-preserving
  node embedding.
\newblock In {\em CIKM}, 2020.

\bibitem{dana}
Tyler Derr, Hamid Karimi, Xiaorui Liu, Jiejun Xu, and Jiliang Tang.
\newblock Deep adversarial network alignment.
\newblock {\em arXiv preprint arXiv:1902.10307}, 2019.

\bibitem{douglas2018metrics}
Joel Douglas, Ben Zimmerman, Alexei Kopylov, Jiejun Xu, Daniel Sussman, and
  Vince Lyzinski.
\newblock Metrics for evaluating network alignment.
\newblock {\em GTA3 at WSDM}, 2018.

\bibitem{gntk}
Simon~S Du, Kangcheng Hou, Russ~R Salakhutdinov, Barnabas Poczos, Ruosong Wang,
  and Keyulu Xu.
\newblock Graph neural tangent kernel: Fusing graph neural networks with graph
  kernels.
\newblock In {\em NeurIPS}, 2019.

\bibitem{du2019joint}
Xingbo Du, Junchi Yan, and Hongyuan Zha.
\newblock Joint link prediction and network alignment via cross-graph
  embedding.
\newblock In {\em IJCAI}, 2019.

\bibitem{dgmc}
Matthias Fey, Jan~E Lenssen, Christopher Morris, Jonathan Masci, and Nils~M
  Kriege.
\newblock Deep graph matching consensus.
\newblock In {\em ICLR}, 2020.

\bibitem{gold1996graduated}
Steven Gold and Anand Rangarajan.
\newblock A graduated assignment algorithm for graph matching.
\newblock {\em TPAMI}, 1996.

\bibitem{hashalign}
Mark Heimann, Wei Lee, Shengjie Pan, Kuan-Yu Chen, and Danai Koutra.
\newblock Hashalign: Hash-based alignment of multiple graphs.
\newblock In {\em PAKDD}, 2018.

\bibitem{rgm}
Mark Heimann, Tara Safavi, and Danai Koutra.
\newblock Distribution of node embeddings as multiresolution features for
  graphs.
\newblock In {\em ICDM}, 2019.

\bibitem{regal}
Mark Heimann, Haoming Shen, Tara Safavi, and Danai Koutra.
\newblock Regal: Representation learning-based graph alignment.
\newblock In {\em CIKM}, 2018.

\bibitem{bigalign}
Danai Koutra, Hanghang Tong, and David Lubensky.
\newblock Big-align: Fast bipartite graph alignment.
\newblock In {\em ICDM}, 2013.

\bibitem{graal}
Oleksii Kuchaiev, Tijana Milenkovi{\'c}, Vesna Memi{\v{s}}evi{\'c}, Wayne
  Hayes, and Nata{\v{s}}a Pr{\v{z}}ulj.
\newblock Topological network alignment uncovers biological function and
  phylogeny.
\newblock {\em Journal of the Royal Society Interface}, 2010.

\bibitem{koblenz}
J{\'e}r{\^o}me Kunegis.
\newblock Konect: the koblenz network collection.
\newblock In {\em WWW}. ACM, 2013.

\bibitem{lample2018word}
Guillaume Lample, Alexis Conneau, Marc'Aurelio Ranzato, Ludovic Denoyer, and
  Hervé Jégou.
\newblock Word translation without parallel data.
\newblock In {\em ICLR}, 2018.

\bibitem{snapnets}
Jure Leskovec and Andrej Krevl.
\newblock {SNAP Datasets}: {Stanford} large network dataset collection.
\newblock \url{http://snap.stanford.edu/data}, June 2014.

\bibitem{li2019adversarial}
Chaozhuo Li, Senzhang Wang, Yukun Wang, Philip Yu, Yanbo Liang, Yun Liu, and
  Zhoujun Li.
\newblock Adversarial learning for weakly-supervised social network alignment.
\newblock In {\em AAAI}, 2019.

\bibitem{liu2016aligning}
Li~Liu, William~K Cheung, Xin Li, and Lejian Liao.
\newblock Aligning users across social networks using network embedding.
\newblock In {\em IJCAI}, 2016.

\bibitem{melnik2002similarity}
Sergey Melnik, Hector Garcia-Molina, and Erhard Rahm.
\newblock Similarity flooding: A versatile graph matching algorithm and its
  application to schema matching.
\newblock In {\em ICDE}, 2002.

\bibitem{meng2016igloo}
Lei Meng, Joseph Crawford, Aaron Striegel, and Tijana Milenkovic.
\newblock Igloo: Integrating global and local biological network alignment.
\newblock In {\em MLG at KDD}, 2016.

\bibitem{patro2012global}
Rob Patro and Carl Kingsford.
\newblock Global network alignment using multiscale spectral signatures.
\newblock {\em Bioinformatics}, 2012.

\bibitem{qin2020g}
Kyle~K Qin, Flora~D Salim, Yongli Ren, Wei Shao, Mark Heimann, and Danai
  Koutra.
\newblock G-crewe: Graph compression with embedding for network alignment.
\newblock In {\em CIKM}, 2020.

\bibitem{magna}
Vikram Saraph and Tijana Milenkovi{\'c}.
\newblock Magna: maximizing accuracy in global network alignment.
\newblock {\em Bioinformatics}, 2014.

\bibitem{isorank}
Rohit Singh, Jinbo Xu, and Bonnie Berger.
\newblock Global alignment of multiple protein interaction networks with
  application to functional orthology detection.
\newblock {\em PNAS}, 2008.

\bibitem{sinkhorn1964relationship}
Richard Sinkhorn.
\newblock A relationship between arbitrary positive matrices and doubly
  stochastic matrices.
\newblock {\em Ann. Math. Stat.}, 1964.

\bibitem{wang2020credible}
Chenxu Wang, Yang Wang, Zhiyuan Zhao, Dong Qin, Xiapu Luo, and Tao Qin.
\newblock Credible seed identification for large-scale structural network
  alignment.
\newblock {\em DAMI}, 2020.

\bibitem{sgc}
Felix Wu, Amauri Souza, Tianyi Zhang, Christopher Fifty, Tao Yu, and Kilian
  Weinberger.
\newblock Simplifying graph convolutional networks.
\newblock In {\em ICML}, 2019.

\bibitem{wu2020neighborhood}
Yuting Wu, Xiao Liu, Yansong Feng, Zheng Wang, and Dongyan Zhao.
\newblock Neighborhood matching network for entity alignment.
\newblock In {\em ACL}, 2020.

\bibitem{xu2019gromov}
Hongteng Xu, Dixin Luo, Hongyuan Zha, and Lawrence~Carin Duke.
\newblock Gromov-wasserstein learning for graph matching and node embedding.
\newblock In {\em ICML}, 2019.

\bibitem{final}
Si~Zhang and Hanghang Tong.
\newblock Final: Fast attributed network alignment.
\newblock In {\em KDD}, 2016.

\bibitem{zhang2017ineat}
Si~Zhang, Hanghang Tong, Jie Tang, Jiejun Xu, and Wei Fan.
\newblock ineat: Incomplete network alignment.
\newblock In {\em ICDM}, 2017.

\bibitem{zhang2019origin}
Si~Zhang, Hanghang Tong, Jiejun Xu, Yifan Hu, and Ross Maciejewski.
\newblock Origin: Non-rigid network alignment.
\newblock In {\em Big Data}, 2019.

\bibitem{zhang2019kergm}
Zhen Zhang, Yijian Xiang, Lingfei Wu, Bing Xue, and Arye Nehorai.
\newblock {KerGM: Kernelized Graph Matching}.
\newblock In {\em NeurIPS}, 2019.

\end{thebibliography}

\newpage
\appendix
\section{Proofs of Theorems in \S~\ref{sec:theory}-\ref{sec:method}}
\label{app:proofs}

\begin{proof}
(\textbf{Theorem~\ref{thm:mnc-noisy}: Perfect alignment accuracy implies high MNC}).
By Eq.~\eqref{eq:jaccard}, {\small $\text{MNC}(i,\pi(i)) = \frac{|\tilde{\mathcal{N}}_{\overline{\graph}_2}^\pi(i) \cap \mathcal{N}_{\overline{\graph}_2}(\pi(i))|}{|\tilde{\mathcal{N}}_{\overline{\graph}_2}^\pi(i) \cup \mathcal{N}_{\overline{\graph}_2}(\pi(i))|}$}.
 By definition, {\small $\tilde{\mathcal{N}}_{\graph_2}^\pi(i) = \tilde{\mathcal{N}}_{\overline{\graph}_2}^\pi(i)$}; it does not change as neither $\pi$ nor $\graph_1$'s adjacency matrix $\adj_1$ is affected by the noise.  However, {\small $\mathcal{N}_{\overline{\graph}_2}(\pi(i)) \subseteq \mathcal{N}_{\graph_2}(\pi(i))$}, since under edge removal $\pi(i)$ can only lose neighbors in {\small $\overline{\graph}_2$} compared to {\small $\graph_2$}. 
 
 Now {\small $\tilde{\mathcal{N}}_{\graph_2}^\pi(i) = \mathcal{N}_{\graph_2}(\pi(i))$} since by definition an isomorphism is edge preserving, 
 and so {\small $\mathcal{N}_{\graph_2}(\pi(i)) = \tilde{\mathcal{N}}_{\graph_2}^\pi(i)$}, 
 which is the same as {\small $\tilde{\mathcal{N}}_{\overline{\graph}_2}^\pi(i)$}. 
 Thus, {\small $\mathcal{N}_{\overline{\graph}_2}(\pi(i)) \subseteq \tilde{\mathcal{N}}_{\overline{\graph}_2}^\pi(i)$}.  
 We can simplify 
 {\small $\text{MNC}(i,\pi(i)) = \frac{|\mathcal{N}_{\overline{\graph}_2}(\pi(i))|}{|\tilde{\mathcal{N}}_{\overline{\graph}_2}^\pi(i)|} = \frac{|\mathcal{N}_{\overline{\graph}_2}(\pi(i))|}{|\mathcal{N}_{\graph_2}(\pi(i))|}.$} 
 However, every node {\small $j' \in \mathcal{N}_{\graph_2}(\pi(i))$} is also in {\small $\mathcal{N}_{\overline{\graph}_2}(\pi(i))$} as long as the edge {\small $\big(\pi(i),j')\big) \in \edgeSet_2$} has not been removed from {\small $\tilde{\edgeSet}_2$}, which happens with probability $p$. So, {\small $\mathbb{E}\Big( \frac{|\mathcal{N}_{\overline{\graph}_2}(\pi(i))|}{|\mathcal{N}_{\graph_2}(\pi(i))|} \Big) = \mathbb{E}\big(\text{MNC}(i,\pi(i))\big) = 1 - p$}.
\end{proof}

\begin{proof}
(\textbf{Theorem~\ref{thm:mnc-acc}: Perfect MNC implies perfect accuracy except for structurally indistinguishable nodes}).  Since for isomorphic graphs, a node $v$ is structurally indistinguishable from its true counterpart $v'$, and since graph isomorphism is transitive, it suffices to show that $v^*$ is also structurally indistinguishable from $v$.
Suppose for some $k$, $\mathcal{N}_k(v)$ is not isomorphic to $\mathcal{N}_k(v^*)$.  Then by definition there exists neighboring nodes $a,b \in \mathcal{N}_k(v)$ where either $\pi(a)$ or $\pi(b)$ is not in $\mathcal{N}_k(v^*)$, or $\pi(a)$ and $\pi(b)$ do not share an edge.  

In case 1, without loss of generality $\pi(b) \notin \mathcal{N}_k(v^*)$.  Then 
no bijective mapping exists between a shortest path between $v^*$ and $\pi(b)$ and a shortest path from $v^*$ to $\pi(b)$.  There will thus be neighbors on one path whose counterparts are not neighbors on the other path, making MNC less than 1: a contradiction. 

In case 2, since $\pi(b)$ is the counterpart of a neighbor of $a$, it must also be a neighbor of the counterpart of $a$, which is a contradiction of the assumption that $\pi(a)$ and $\pi(b)$ do not share an edge, or else MNC($a,\pi(a)$) $<$ 1, another contradiction.  Thus, we conclude that the $k$-hop neighborhoods are isomorphic, that is, $v$ and $v^*$ are structurally indistinguishable.
\end{proof}

\begin{proof}
(\textbf{Theorem \ref{thm:update}: Matrix form of MNC}).
{
{\small $\tilde{\mathcal{N}}_{\graph_2}^\pi(i) = \{\ell: \exists k \in \vertexSet_1 \text{ s.t. } \adj_{1_{ik}} \matchMat_{k\ell} \neq 0 \}$}, and of course {\small $\mathcal{N}_{\graph_2}(j) = \{\ell : \adj_{2_{j\ell}} \neq 0\}$}.  Since the product of two numbers is nonzero if and only if both numbers are nonzero, {\small $\tilde{\mathcal{N}}_{\graph_2}^\pi(i) \cap \mathcal{N}_{\graph_2}(j) = \{\ell : \adj_{1_{ik}} \matchMat_{k\ell} \adj_{2_{j\ell}} \neq 0 \}$}.  For binary {\small $\adj_1, \adj_2,$} and {\small $\matchMat$}, the cardinality of this set, which is the numerator of Eq.~\eqref{eq:jaccard}, is {\small $\sum_{k \in \vertexSet_1, \ell \in \vertexSet_2}\adj_{1_{ik}} \matchMat_{k\ell} \adj_{2_{j\ell}} = (\adj_1 \matchMat \adj_2)_{ij}$}.   
\noindent Meanwhile, the denominator of Eq.~\eqref{eq:jaccard} is {\small $|\tilde{\mathcal{N}}_{\graph_2}^\pi(i) \cup \mathcal{N}_{\graph_2}(j)| = |\tilde{\mathcal{N}}_{\graph_2}^\pi(i)| + |\mathcal{N}_{\graph_2}(j)| - |\tilde{\mathcal{N}}_{\graph_2}^\pi(i) \cap \mathcal{N}_{\graph_2}(j)|$}. 
Plugging in for each individual term, we obtain {\small $\sum_{k \in \vertexSet_1}\sum_{\ell \in \vertexSet_2} \adj_{1_{ik}} \matchMat_{k\ell} + \sum_{\ell \in \vertexSet_2} \adj_{2_{j \ell}} - \sum_{k \in 
\vertexSet_1}\sum_{\ell \in \vertexSet_2}\adj_{1_{ik}} \matchMat_{k\ell} \adj_{2_{j\ell}}.$} 
Substituting matrix products, this becomes {\small $\sum_{\ell \in \vertexSet_2} (\adj_1\matchMat)_{i\ell} + \sum_{\ell \in \vertexSet_2} \adj_{2_{j\ell}} - (\adj_1 \matchMat \adj_2)_{ij}$}. 
Using all-$1$ vectors to sum over columns, this is
{\small $(\adj_1\matchMat \mathbf{1}^{n_2})_i + (\adj_2 \mathbf{1}^{n_2})_{j} - (\adj_1 \matchMat \adj_2)_{ij}.$} 
Then, expanding the two left vectors into matrices with outer product: 
{\small $(\adj_1\matchMat \mathbf{1}^{n_2} \otimes \mathbf{1}^{n_2})_{ij} + (\textbf{1}^{n_1} \otimes \adj_2 \mathbf{1}^{n_2})_{ij} - (\adj_1 \matchMat \adj_2)_{ij}.$} 
Adding everything together, the denominator is the $ij$-th entry of the matrix 
{\small $\adj_1\matchMat \mathbf{1}^{n_2} \otimes \mathbf{1}^{n_2} + \textbf{1}^{n_1} \otimes \adj_2 \mathbf{1}^{n_2} - \adj_1 \matchMat \adj_2$}.  
}
\end{proof}

\section{Connections to Other Graph Methods}
\label{app:connections}

We show additional connections between \method and other diverse graph methods: first, \textit{seed-and-extend} as an alignment strategy, and second a graph filtering perspective on \method's update rule similar to the analysis of graph neural networks. 

\vspace{0.1cm}
\noindent \textbf{Seed-and-extend alignment heuristic}. Many global network alignment methods~\cite{graal,patro2012global} use this heuristic to find node correspondences between two or multiple networks. Given initial pairwise similarities (or alignment costs) between nodes of the compared graphs  as  $\matchMat$, a pair of nodes $i$ and $j$ with high probability to be aligned (e.g., whose similarity according to $\matchMat$ is above some confidence threshold) are set as the seed regions of the alignment.
  After the seed $(i,j)$ is selected, the $r\textendash$hop neighborhoods of $i$ and $j$ (i.e., $\mathcal{N}_{r, \graph_1}(i)$ and $\mathcal{N}_{r, \graph_2}(j)$) 
  in their respective graphs are built. Next, the selected seed $(i,j)$ is extended for the final alignment $\matchMat'$ by greedily matching nodes in $\mathcal{N}_{r,\graph_1}(i)$ and $\mathcal{N}_{r, \graph_2}(j)$, 
  searching for the pairs $(i',j'):i' \in \mathcal{N}_{r,\graph_1}(i)$ and $j' \in \mathcal{N}_{r,\graph_2}(j)$ that are not already aligned and can be aligned with the maximum value of similarity according to $\matchMat$.
  The process can be written as $\forall k \in \mathcal{N}_{r,\graph_1}(i), \matchMat'_{k\ell} \neq 0 \text{ if } \ell = \arg\max_{\ell \in \mathcal{N}_{r,\graph_2}(j)} \matchMat_{k\ell}$.

By setting $r = 1$ to consider each seed's direct neighbors, and $\matchMat'_{k^*\ell^*} \neq 0$ if $k^*,\ell^* = \arg \max_{k \in \vertexSet_1, \ell \in \vertexSet_2} \adj_{1_{ik}} \matchMat_{k\ell} \adj_{2_{j\ell}}$, 
we see that the seed-and-extend heuristic analyzes the same set of elements used to compute the update in \method (Eq.~\eqref{eq:update}). However, instead of summing them to update the similarity of seed nodes $i$ and $j$, it takes the argmax over them to adaptively select the next pair of alignments.  Thus, seed-and-extend aligns less well with \textbf{\insightone} by relying heavily on a correct initial solution, as the early alignments are irrevocable and used to restrict the scope of subsequent alignments.  

\vspace{0.1cm}
\noindent  \textbf{Graph Filtering.}  The matching matrix $\matchMat$ can also be interpreted as a high-dimensional feature matrix.  For example, for each node in $\graph_1$, a row of $\matchMat$ may be regarded as an $\numberOfNodes{2}$-dimensional feature vector consisting of the node correspondences to each of the nodes in $\graph_2$, and similarly the $\numberOfNodes{2} \times \numberOfNodes{1}$ matrix $\matchMat^\top$
contains $\numberOfNodes{1}$-dimensional cross-network correspondence features for each node in $\graph_2$.  
For challenging alignment scenarios, these are likely highly noisy features.  However, recent works have shown that multiplying a node feature matrix by the graph's adjacency matrix corresponds to a low-pass filtering operation, which is of interest in explaining the mechanisms of graph neural networks~\cite{sgc}.  

We can write our update rule in Eq.~\eqref{eq:update} as a feature matrix left multiplied by adjacency matrices: $\adj_1 \matchMat \adj_2 = \adj_1 (\adj_2 \matchMat^\top)^\top$ (for undirected graphs, $\adj_2^\top = \adj_2$), where   $\adj_2 \matchMat^\top$ produces a filtered set of $\numberOfNodes{1}$-dimensional features. By taking the transpose, these may be interpreted as $\numberOfNodes{2}$-dimensional features for each node of $\graph_1$, which are then filtered again by left multiplication with $\adj_1$.\footnote{Graph convolutional networks use the augmented normalized adjacency matrix $\tilde{\mathbf{D}}^{-\frac{1}{2}} \tilde{\adj} \tilde{\mathbf{D}}^{-\frac{1}{2}} \text{ where } \tilde{\adj} = \adj + \mathbf{I}$, which we have not found helpful.}  

Interpreting \method's updates as graph filtering explains its strong performance, as well as the success of recent supervised graph matching work~\cite{zhang2019origin,dgmc} using graph neural networks.  Of course \method does not have the learnable parameters and nonlinearities of a graph neural network.  However, just as SGC~\cite{sgc} recently compares favorably to graph neural networks by replacing their deep nonlinear feature extraction with repeated multiplication by the adjacency matrix, we find that that unsupervised ``alignment filtering'' is highly effective.    

\vspace{0.1cm}
\noindent  \textbf{Graph Kernels.}
Just as nodes can be matched across networks using suitable cross-network node similarities, entire graphs can be compared (for the purposes of performing machine learning tasks on graph-structured data points) by aggregating their nodes' similarities~\cite{rgm,gntk}.  Some base network alignment methods we studied are known to have close connections to graph kernels.  For example, FINAL's pooled cross-network node similarities are closely related to a graph kernel based on random walks~\cite{final}; likewise, the xNetMF node embeddings that REGAL uses to match nodes one by one~\cite{regal} can also be used to construct a feature map for an entire graph, where the dot product between two graphs' feature maps approximates the mean-pooled (kernelized) node embedding similarities~\cite{rgm}.  We now show that the update rule of \method is also related to the graph neural tangent kernel (GNTK) ~\cite{gntk}, a graph kernel designed to achieve the effect of an infinitely wide graph neural network trained by gradient descent.  
\\
GNTK starts by computing a cross-network node similarity matrix, the same as our matrix $\matchMat$.  (As proposed in~\cite{gntk}, this matrix is computed using input node features that are common in graph classification benchmark datasets, but the formulation can be more general.) To achieve the effect of a graph neural network's feature aggregation, the node similarities are iteratively propagated across graphs: 
\\
$$\matchMat'_{ij} = c_i c_j \sum\limits_{u \in \{\mathcal{N}_{\graph_1}(i) \cup i \}} \sum\limits_{v \in \{\mathcal{N}_{\graph_2}(j) \cup j \}} \matchMat_{uv}$$
\\
Up to scaling factors $c_i$ and $c_j$ designed to model the effects of particular graph neural network architectures, this term is elementwise equivalent to our update $\matchMat' = \adj_1 \matchMat \adj_2$, where the graphs' adjacency matrices have been augmented with self-loops as is commonplace for graph neural networks. At each round of propagation, GNTK post-processes $\matchMat$ by performing $R$ transformations corresponding to $R$ fully-connected neural network layers with ReLU activation, whereas we simply add token match scores and normalize.  
\\
The analysis for GNTK connects the training of kernel machines (the final kernel value is obtained by applying a READOUT operation to the refined node similarities, a simple example being to sum them all) with that of graph neural networks, in either case using graph-level supervision~\cite{gntk}.  While this is of course a quite different setup from the unsupervised node-level matching problem \method tackles, a very interesting direction for future work is to see if \method and/or GNTK can benefit from further methodological exchange (at minimum, our sparse refinement techniques could approximate the computation of GNTK between large graphs).  Such analysis may also reveal more about the general connection between network alignment based on individual node-level similarities and network classification based on aggregated node-level similarities.

\section{Complexity Analysis of \method}
\label{app:complexity}
Here, we analyze the precise computational complexity of \method with both sparse and dense refinement.  

\subsection{Complexity of Dense Refinement.} 
To simplify notation, we assume that both graphs have $\numberOfAllNodes$ nodes~\cite{regal}.  
For $K$ iterations, our algorithm computes the left and right multiplication of a dense $\numberOfAllNodes \times \numberOfAllNodes$ matching matrix with two adjacency matrices of graphs with average degree (number of nonzero entries per row of the adjacency matrix) $\degavg_1$ and $\degavg_2$, respectively.  Thus, the time complexity of this update step is $O(\numberOfAllNodes^2(\degavg_1 + \degavg_2))$.  Normalizing the matrix at each iteration and adding in token match scores requires $O(\numberOfAllNodes^2)$ time.  Therefore, the overall time complexity is $O\Big(K \numberOfAllNodes^2 (\degavg_1 + \degavg_2)\Big)$.  While the number of iterations and average node degree are in practice constant or asymptotically smaller than the graph size, the quadratic time and space complexity of \method's dense matrix operations makes it harder to apply to large graphs.

\subsection{Complexity of Sparse Refinement.}
For $K$ iterations, we compute a sparse update matrix with $O(\numberOfAllNodes \alpha)$ nonzero entries by multiplying matrices with $O(\numberOfAllNodes \degavg_1)$, $O(K \numberOfAllNodes \alpha)$, and $O(\numberOfAllNodes \degavg_2)$ nonzero entries, respectively.  It takes $O(\numberOfAllNodes K \alpha \degavg_1)$ time to compute $\tilde{\adj_1} = \adj_1 \matchMat_{k-1}$ and $O(\numberOfAllNodes K \alpha \degavg_1 \degavg_2)$ time to compute $\tilde{\adj_1} \adj_2$.  We then compute $\matchMat_k$ by updating $O(\numberOfAllNodes \alpha)$ entries in $\matchMat_{k-1}$ per iteration.  Thus, $\matchMat_k$ may have $O(K \numberOfAllNodes \alpha)$ nonzero entries and requires $O(K \numberOfAllNodes \alpha)$ time to update and normalize.  Altogether, the runtime is now $O(\numberOfAllNodes K^2 \alpha \degavg_1 \degavg_2)$, i.e. linear in the number of nodes.  (This is a worst-case analysis; in practice, the runtime scales close to linearly with $K$.)  We can also avoid storing a dense matching matrix, leading to subquadratic space complexity.

\section{Baseline Hyperparameter Settings}
\label{app:params}
For REGAL's own xNetMF embeddings, we used default embedding dimension $\lfloor 10 \log_2(\numberOfNodes{1} + \numberOfNodes{2}) \rfloor$~\cite{regal}, maximum neighborhood distance $2$, neighborhood distance discount factor $\delta = 0.1$, and resolution parameter $\gamma_{\text{struc}} = 1$, all recommended parameters.  
For the embeddings used in CONE-Align, we set embedding dimension $d=128$, context window size $w=10$, and negative sampling parameter $\alpha=1$. We used ${n_0} = 10$ iterations and regularization parameter $\reg_0 = 1.0$ for the convex initialization of the subspace alignment, which we performed with $T = 50$ iterations of Wasserstein Procrustes optimization with batch size $b = 10$, learning rate $\lr = 1.0$, and regularization parameter $\reg = 0.05$ as were suggested by the authors~\cite{conealign}.  For REGAL and CONE-Align, we computed embedding similarity with dot product followed by softmax normalization~\cite{dgmc}, using $k$-$d$ trees to perform fast 10-nearest neighbor search for REGAL on LiveMocha~\cite{regal}.  

NetAlign and FINAL require a matrix of prior alignment information, which we computed from pairwise node degree similarity. Then following~\cite{regal,dana}, we constructed this matrix by taking the top $k =\lfloor \log_2 \big(\numberOfNodes{1} + \numberOfNodes{2})/2\big) \rfloor$) entries for each node in $\graph_1$; that is, the top $k$ most similar nodes in $\graph_2$ by degree.

For MAGNA, starting from a random initial population with size 15000, we simulated the evolution for 2000 steps following~\cite{magna} using edge correctness as the optimizing measure as it is similar to the objectives of our other methods.  
We used 3 threads to execute the alignment procedure.

We consider the output of all methods to be a binary matrix $\matchMat$ consisting of the ``hard'' (one-to-one) alignments they find, to treat methods consistently and to show that \method can refine the most general network alignment solution.  It is worth noting that some methods (e.g. REGAL, CONE-Align) can also produce ``soft'' alignments (real-valued node similarity scores) and our formulation is capable of using those.  

\noindent \textbf{Computing Environment.}
We performed all experiments on an Intel(R) Xeon(R) CPU E5-1650 at 3.50GHz, 256GB RAM.

\section{Convergence Analysis: Accuracy \& MNC}
\label{app:convergence}
\begin{figure*}[t!]
	\centering

    \subfloat{
      \includegraphics[width=.85\textwidth]{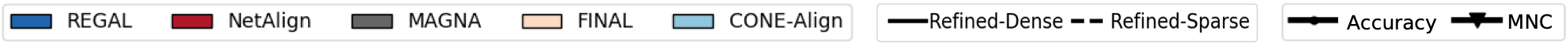}
    }
	
	 \setcounter{subfigure}{0}
    \subfloat[Arenas 5\% noise\label{arenas5}]{
      \includegraphics[width=0.19\textwidth, trim=0 0 1.5cm 1.5cm, clip]{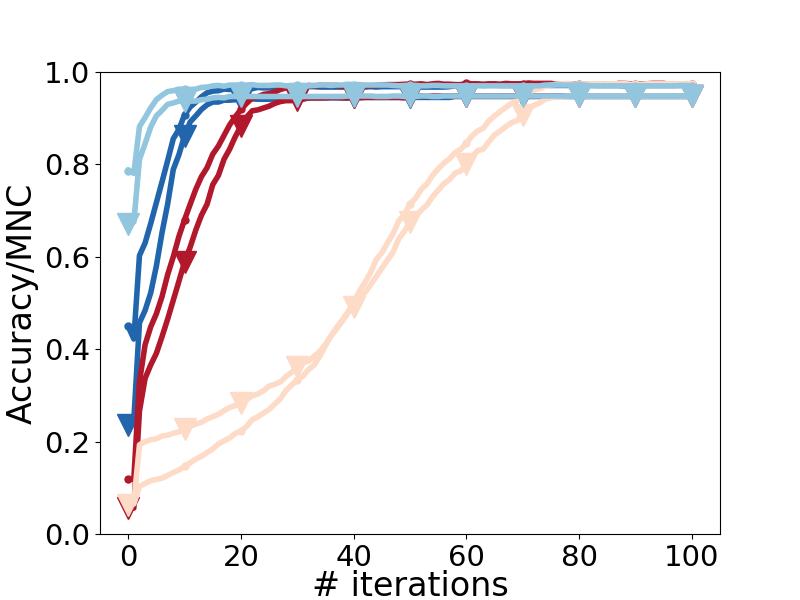}
    }
    \subfloat[Hamsterster 5\% noise\label{hamster5}]{
      \includegraphics[width=0.19\textwidth, trim=0 0 1.5cm 1.5cm, clip]{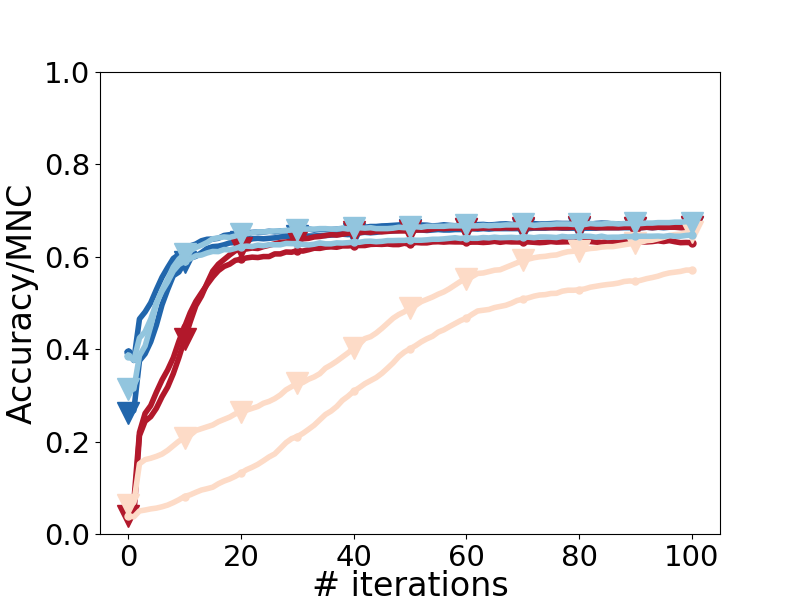}
    }
    \subfloat[PPI-H 5\% noise\label{ppi5}]{
      \includegraphics[width=0.19\textwidth, trim=0 0 1.5cm 1.5cm, clip]{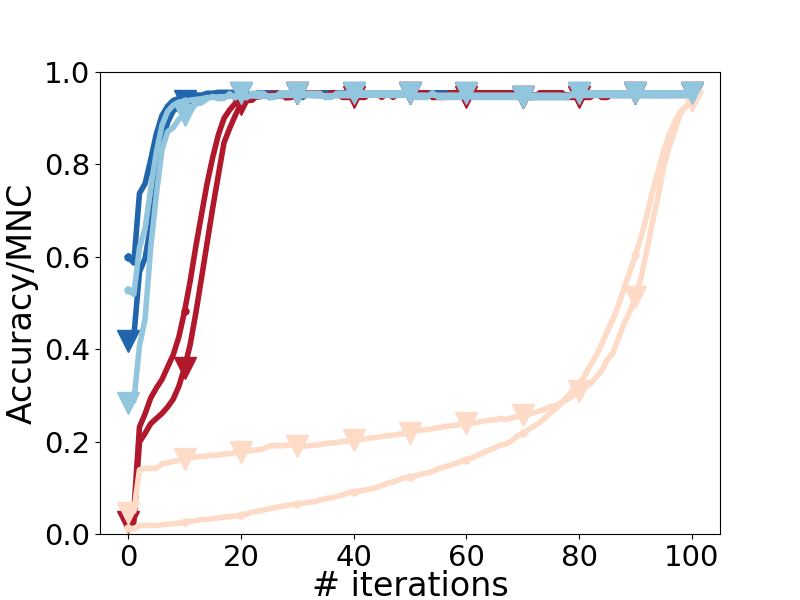}
    }
    \subfloat[Facebook 5\% noise
    \label{fb5}]{\includegraphics[width=0.19\textwidth, trim=0 0 1.5cm 1.5cm, clip]{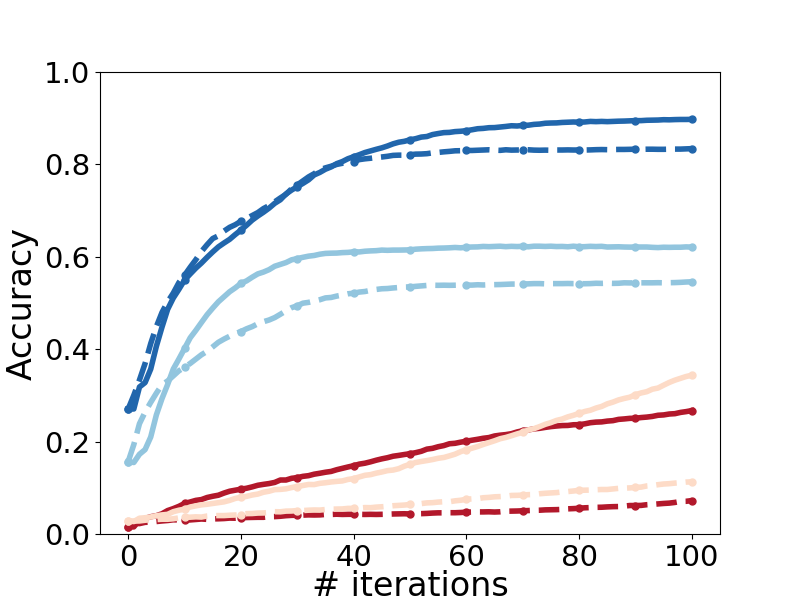}
    }
    \subfloat[PPI-Y 5\% added edges
    \label{magna5}]{\includegraphics[width=0.19\textwidth, trim=0 0 1.5cm 1.5cm, clip]{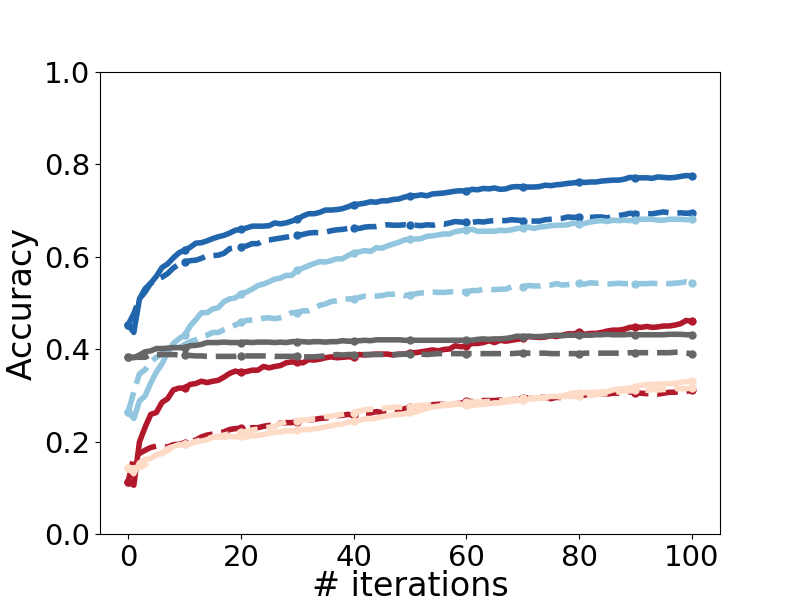}
    }
    
    \subfloat[Arenas 25\% noise\label{arenas25}]{
      \includegraphics[width=0.19\textwidth, trim=0 0 1.5cm 1.5cm, clip]{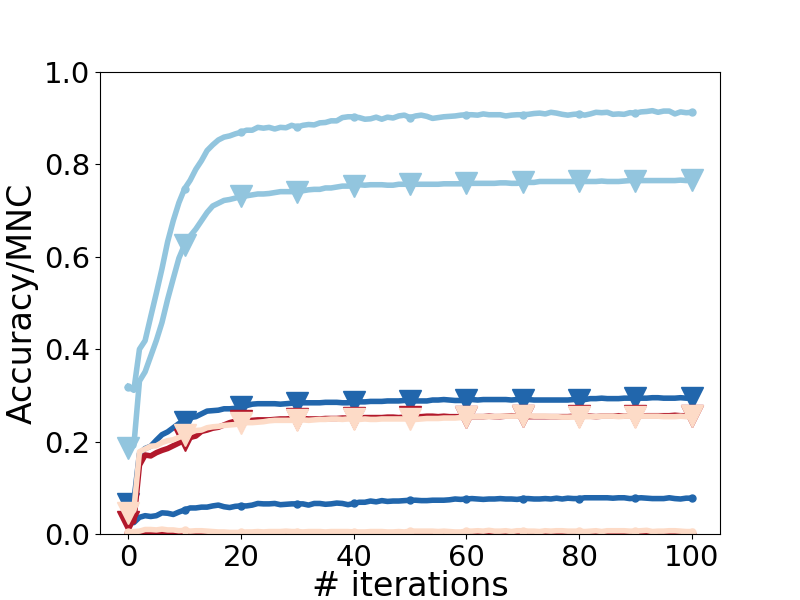}
    }
    \subfloat[Hamsterster 25\% noise\label{hamster25}]{
      \includegraphics[width=0.19\textwidth, trim=0 0 1.5cm 1.5cm, clip]{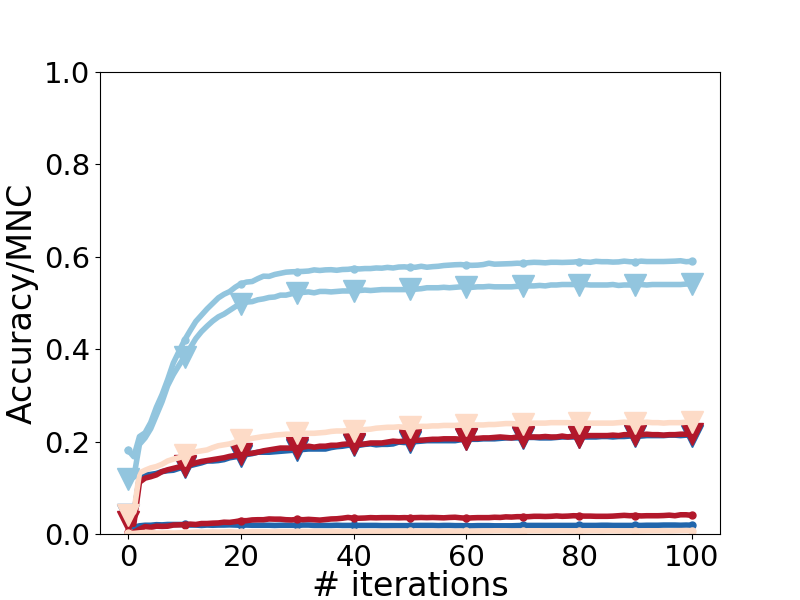}
    }
    \subfloat[PPI-H 25\% noise\label{ppi25}]{
      \includegraphics[width=0.19\textwidth, trim=0 0 1.5cm 1.5cm, clip]{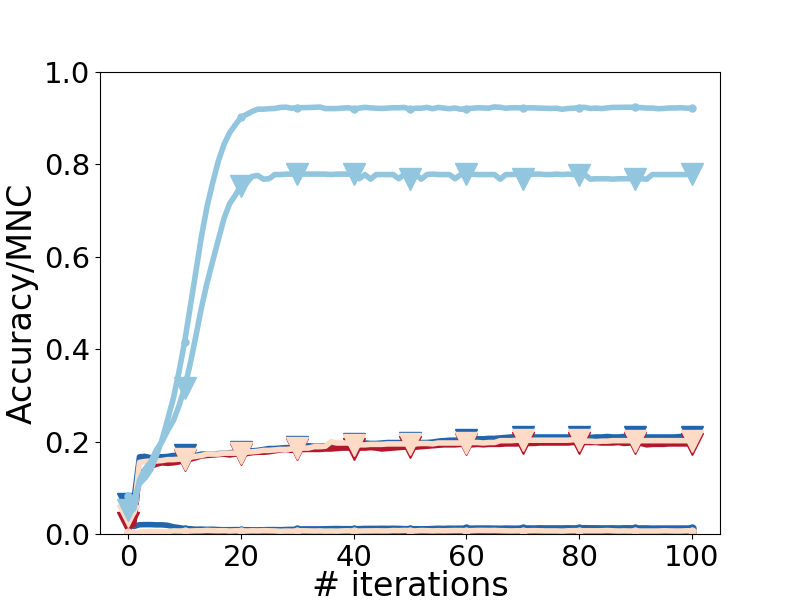}
    }
    \subfloat[Facebook 25\% noise
    \label{fb25}]{\includegraphics[width=0.19\textwidth, trim=0 0 1.5cm 1.5cm, clip]{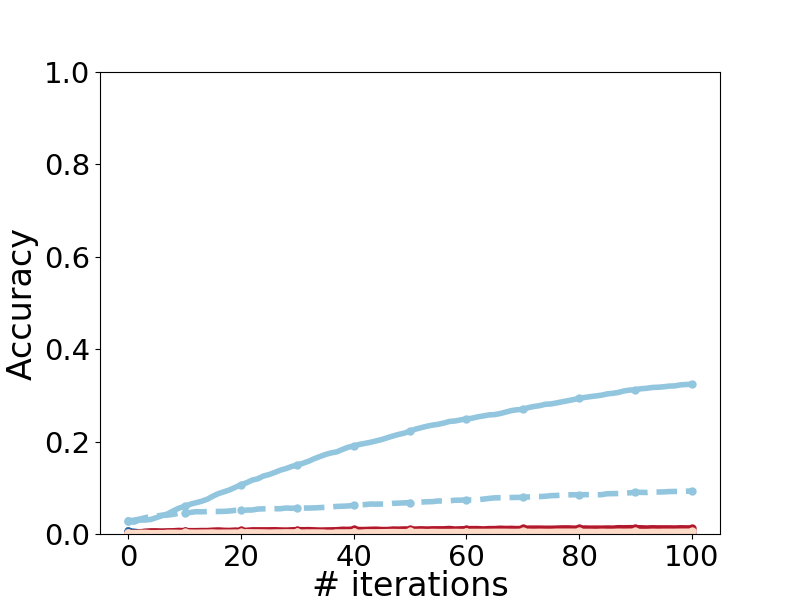}
    }
    \subfloat[PPI-Y 25\% added edges
    \label{magna25}]{\includegraphics[width=0.19\textwidth, trim=0 0 1.5cm 1.5cm, clip]{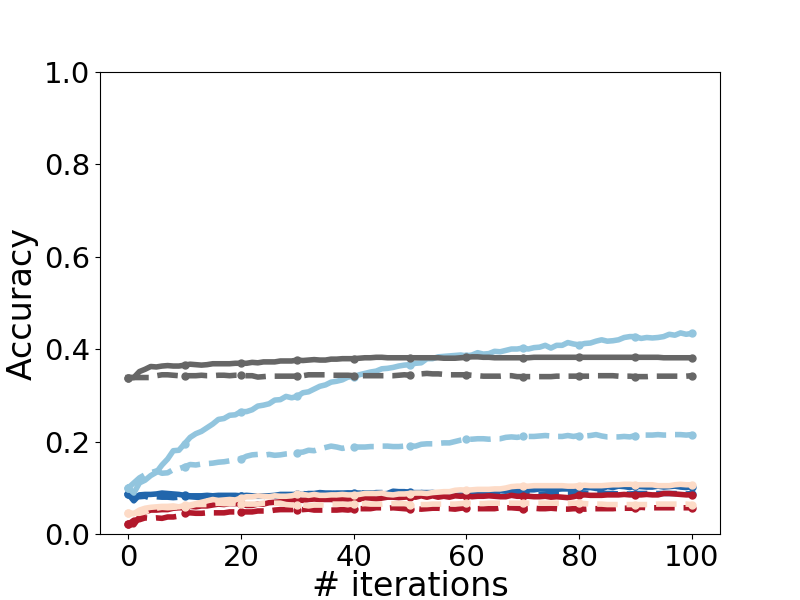}
    }
	\caption{Analysis of \method as a function of number of iterations (0 = performance of base method before refinement).  Arenas, PPI-H, and Hamsterster datasets show accuracy and MNC, and Facebook and PPI-Y datasets show accuracy of sparse and dense \method versions. Convergence rates are different for different methods, but often happen well before 100 iterations for both sparse and dense \method.  Accuracy and MNC consistently increase and follow similar trends, as per their theoretical connection.}
    \label{fig:niter}
\end{figure*}

One of the parameters of \method is $K$, the number of iterations for which the initial matching is refined.  
In Fig.~\ref{fig:niter}, we plot the performance at each iteration, up to our maximum value of 100, for all methods and datasets (at lowest and highest noise levels, or number of added low-confidence PPIs in the case of PPI-Y). 
For brevity, we show accuracy and MNC for dense refinement only on the first three datasets (Arenas, Hamsterster, and PPI-H), and the per-iteration accuracy only for sparse and dense refinement on the remaining datasets.

\vspace{0.1cm}
\noindent \textit{Results.} We see that accuracy and MNC tend to trend similarly, as do sparse and dense refinement. In both cases, we see similar trends for the same colored curves. As for \method variations, dense refinement of CONE-Align grows in accuracy slightly more steeply than sparse refinement.  For MNC, we see that accuracy and MNC tend to have very similar values at 5\% noise (thus the lines of the same color are close to overlapping). At 25\% noise, MNC is lower than accuracy for highly accurate methods like CONE-Align---this is to be expected because Theorem~\ref{thm:mnc-noisy} proved that the expected average MNC under even for a perfect alignment solution is bounded by the noise ratio.  For the remaining methods which struggle to achieve high accuracy, MNC is higher than accuracy.  As Theorem~\ref{thm:mnc-acc} showed, it is possible to find a high-MNC alignment that is still not perfectly accurate (due to structural indistinguishability).  Indeed, here we see this happening in some especially challenging settings starting from less favorable initial solutions.  Nevertheless, in most cases, improving MNC is a successful strategy for accurate network alignment.

Convergence rates differ for different methods and datasets.  For example, FINAL is generally the slowest method to converge, taking close to 70 iterations to fully converge on the Arenas dataset, with NetAlign taking around 20, REGAL around 10, and CONE-Align around 5.  On the PPI-H dataset with 5\% noise, FINAL sees slow progress for nearly the full 100 iterations before making rapid progress at the end.  

In general, we find that a modest number of iterations is generally sufficient.  While we allow all methods 100 iterations of refinement in our experiments, the elbow-shaped curves indicate that in practice, convergence often happens in far fewer than 100 iterations (with some exceptions for a few methods on the PPI-Y and Facebook datasets).  In practice, early convergence can be ascertained by 
small changes in the discovered alignments.

\begin{observation}
Accuracy and MNC, sparse and dense refinement tend to trend similarly on each method on each dataset.  Although convergence rates differ for different methods and datasets, convergence is quite fast in practice.  
\end{observation}

\section{Improvement thanks to \method}
\label{app:table}
To further illustrate that different methods benefit differently from refinement, we summarize our results from Fig.~\ref{fig:simulated-acc} in tabular format in Tab.~\ref{tab:results}, where we show the maximum improvement in mean accuracy thanks to \method for each base method on each dataset (across refinement types and noise levels, averaged over five trials).  We also show the maximum noise level at which \method is able to yield a noticeable improvement (3\% or more).  As discussed in \S\ref{sec:experiments}, while all network alignment methods benefit from \method, this does not eliminate the effect of the initial solution.  We have also observed that \method poorly refines a randomly initialized alignment matrix on our datasets. Again, this indicates that our contributions complement rather than replace existing network alignment solutions.
\newcommand\mytop{\rule{0pt}{4ex}}
\setlength{\tabcolsep}{1pt}
\begin{table}[ht]
\caption{Maximum improvement thanks to \method for each base method on each dataset (with abbreviated name), and maximum noise level at which \method brings noticeable improvement.  For all methods and all datasets, \method brings dramatic increases in accuracy and can often bring significant improvement even at very high noise levels.}
\label{tab:results}
\centering
{\scriptsize
\begin{tabular}{l ccccc} 
\toprule
& \multicolumn{1}{c}{\bf REGAL} & \multicolumn{1}{c}{\bf NetAlign} & \multicolumn{1}{c}{\bf FINAL} & \multicolumn{1}{c}{\bf CONE-Align} & \multicolumn{1}{c}{\bf MAGNA}  
 \\ \midrule 
{\bf AR} & 
  \begin{tabular}{c} 82.18\% \\\textit{20\% noise}\end{tabular}
& \begin{tabular}{c}84.63\% \\\textit{10\% noise}\end{tabular}
& \begin{tabular}{c}86.93\% \\\textit{5\% noise} \end{tabular}
& \begin{tabular}{c}58.02\% \\\textit{25\% noise} \end{tabular}
& \begin{tabular}{c} N/A \end{tabular}
 \\ \hline
\mytop {\bf P-H} &
\begin{tabular}{c}80.53\% \\\textit{20\% noise}\end{tabular}
& \begin{tabular}{c}90.02\% \\\textit{10\% noise}\end{tabular}
& \begin{tabular}{c}79.57\% \\\textit{5\% noise}\end{tabular}
& \begin{tabular}{c}83.28\% \\\textit{25\% noise}\end{tabular}
& \begin{tabular}{c} N/A \end{tabular}
 \\ \hline
\mytop {\bf HA} & 
\begin{tabular}{c}47.86\% \\\textit{20\% noise}\end{tabular}
& \begin{tabular}{c}55.85\% \\\textit{15\% noise}\end{tabular}
& \begin{tabular}{c}52.68\% \\\textit{10\% noise}\end{tabular}
& \begin{tabular}{c}39.50\% \\\textit{25\% noise}\end{tabular}
& \begin{tabular}{c} N/A \end{tabular}
\\ \hline
\mytop {\bf FB} & 
 \begin{tabular}{c}61.22\% \\\textit{15\% noise}\end{tabular}
& \begin{tabular}{c}23.76\% \\\textit{15\% noise}\end{tabular}
& \begin{tabular}{c}17.03\% \\\textit{5\% noise}\end{tabular}
& \begin{tabular}{c}48.51\% \\\textit{25\% noise}\end{tabular}
& \begin{tabular}{c} N/A \end{tabular}
  \\ \hline
\mytop {\bf P-Y} & 
\begin{tabular}{c}32.17\% \\\textit{20\% noise}\end{tabular}
& \begin{tabular}{c}34.67\% \\\textit{25\% noise}\end{tabular}
& \begin{tabular}{c}18.83\% \\\textit{25\% noise}\end{tabular}
& \begin{tabular}{c}45.72\% \\\textit{25\% noise}\end{tabular}
& \begin{tabular}{c}6.77\% \\\textit{25\% noise}\end{tabular}
 \\
\bottomrule
\end{tabular}
}
\end{table}

\end{document}